\documentclass[11pt]{article}
\usepackage{amssymb,amsmath,amsfonts}
\usepackage{graphicx}
\usepackage{lscape}
\usepackage[dvips]{color}
\usepackage[round]{natbib}
\usepackage{float}
\usepackage[colorlinks=true, urlcolor=blue, citecolor=blue, breaklinks]{hyperref}

\newcommand{\blue}{\color[rgb]{0,0,1}}

\numberwithin{equation}{section}
\numberwithin{figure}{section}
\numberwithin{table}{section}

\allowdisplaybreaks

\begin{document}   

\baselineskip 5mm

\thispagestyle{empty}

\begin{center}
{\LARGE Quantile and Log-Quantile Least Squares 
for Robust-Efficient 
\\[1ex]
Fitting and Validation of Log-Location-Scale Loss Models}

\vspace{15mm}

{\large\sc
Mohammed Adjieteh\footnote[1]{~Mohammed Adjieteh, Ph.D., ASA, is 
an Assistant Professor in the Department of Mathematical Sciences, 
Appalachian State University, Boone, NC 28608, USA. ~~ {\em e-mail\/}: 
~{\blue\tt adjietehma@appstate.edu}}}  

\vspace{1mm}

{\large\em Appalachian State University}

\vspace{10mm}

{\large\sc  
Vytaras Brazauskas\footnote[2]{
~{\sc Corresponding Author}: Vytaras Brazauskas, Ph.D., ASA,
is a Professor in the Department of Mathematical Sciences,  
University of Wisconsin-Milwaukee, P.O. Box 413, Milwaukee, 
WI 53201, USA. ~~ {\em e-mail\/}: ~{\blue\tt vytaras@uwm.edu}}}  

\vspace{1mm}

{\large\em University of Wisconsin-Milwaukee}

\vspace{10mm}

{\footnotesize To appear in {\em Variance}}
\\
{\scriptsize ( {\em Submitted\/}: ~February 23, 2026 
\qquad
{\em Revised\/}: ~June 7, 2026
\qquad
{\em Accepted\/}: ~August 22, 2026 )}

\end{center}

\vspace{3mm}

\begin{quote}
{\bf\em Abstract\/}.
~A variety of models for insurance and other types of losses are 
special cases of the {\em log-location-scale\/} family, with the 
lognormal and Pareto-$I$ distributions being the most prominent 
examples. The latter also serves as a primary example of infinite-mean 
models that often present challenges in risk management. In this paper, 
we utilize two {\em asymptotic\/} theorems -- the joint normality of 
sample quantiles (of {\em i.i.d.\/} random variables) and the delta 
method -- to construct nonlinear and linear regression frameworks 
for estimation and validation of log-location-scale distributions. 
Within these regression frameworks, four equally robust estimators -- 
ordinary and generalized quantile (oQLS and gQLS) and log-quantile 
(log-oQLS and log-gQLS) least squares -- are proposed. For 
log-location-scale loss models, the logarithmic transformation of 
quantiles approximates the nonlinear least squares solution 
{\em exactly\/} and yields more accurate estimators. Also, the 
log-linear regression framework facilitates a convenient way 
to study the estimators' properties and to design a residuals-based 
goodness-of-fit test. Moreover, log-oQLS and log-gQLS have explicit 
formulas and can be easily computed for medium- ($n=10^3$), large- 
($n=10^4$), and very large-size ($n > 10^6$) samples. Computational 
and statistical performances of the estimators, outlier-labeling 
rules, and the goodness-of-fit test are illustrated using simulated 
and real datasets. 

\vspace{2mm}


\end{quote}

\vspace{4mm}

\noindent
{\bf\em Dedication\/}. ~This paper is dedicated to the memory of 
Professor Ri{\v{c}}ardas Zitikis -- a dear friend, an inspiring
mentor, and a brilliant researcher -- who over the years has 
shared many insights about the effectiveness of quantiles in 
Economics, Finance, Quantitative Risk Management, and other 
fields.

\newpage

\baselineskip 7mm

\setcounter{page}{1}

\section{Introduction}

The claim severity models help insurance companies assess risk, set 
appropriate premiums, and ensure they have enough reserves to cover 
potential claims. Given that such models cannot be constructed without 
making assumptions, which may be reasonable but not perfectly correct, 
their estimation has to be resistant to various potential violations 
of such assumptions. In other words, the model fitting techniques and 
validation tools must help actuaries to manage `model risk'. Among 
many manifestations of model risk, outliers seem to be the most 
influential, evasive, and can be quite harmful in ensuing applications. 
Therefore, robust-efficient fitting of these models is crucial because 
it yields accurate estimates of risk, even in the presence of outliers 
or non-ideal data.

The literature on robust statistics is well established and offers 
a rich variety of techniques for robust-efficient fitting of 
parametric models; they are summarized in the classic books of
\citet[][]{hrrs86},
\citet[][]{mmy06},
and \citet[][]{hr09}.
In actuarial science, robustness studies have been carried out by 
many authors who relied upon varying interpretations of robust 
models. For example, using the traditional interpretation (as 
defined in the aforementioned books), robustification of 
{\em credibility\/} models was pursued by
\citet[][]{k92}, 
\citet[][]{gr93},
\citet[][]{db07},
\citet[][]{kj13},
and
\citet[][]{zp24}.
Further, robust-efficient estimation of {\em heavy-tailed models\/} 
was explored by
\citet[][]{bs00},
\citet[][]{s02b},
\citet[][]{bjz09},
\citet[][]{zbg18},
and
\citet[][]{f22}.
Extensions of robust methods to {\em insurance payment data\/} 
(e.g., left-truncated and right-censored data) have been 
initiated by 
\citet[][]{p21a}.
Finally, other interpretations of robustness have been used in 
actuarial research as well. For example, adopting the minimum 
distance approach (i.e., the distance between probability 
distributions as presented in the book by \citet[][]{bsp11}),
{\em robust risk analysis\/} of several actuarial problems was 
carried out by \citet[][]{blty19}. Yet another approach is to 
incorporate distributional robustness by investigating the 
{\em worst- and best-case\/} scenarios for risk measures. 
Recent examples of this line of work include \citet{lm22} 
and \citet{bpv24}, among many others. The methodology 
proposed in this paper relies upon the traditional 
interpretation of robustness.

A variety of loss severity distributions are special cases of 
the {\em log-location-scale\/} family, with the lognormal and 
Pareto-$I$ distributions being the most prominent examples. 
The latter distribution also serves as a primary example of 
infinite-mean models that often present challenges in risk 
management; see, for example, \citet[][]{cw25} for an extensive
review. In this paper, we build upon the work of \citet{ab25} 
for location-scale families and construct nonlinear and linear 
regression frameworks for estimation and validation of 
log-location-scale distributions, which is accomplished by 
utilizing two {\em asymptotic\/} theorems -- the joint normality 
of sample quantiles (of {\em i.i.d.\/} random variables) and 
the delta method. Within these frameworks, four equally robust 
estimators -- ordinary and generalized quantile (oQLS and gQLS) 
and log-quantile (log-oQLS and log-gQLS) least squares -- are 
introduced. The novelty of the proposed methodology lies in 
the exploitation of the asymptotic structure of the covariance 
matrix of sample quantiles, which is used to construct more 
accurate estimators. The new estimators offer attractive 
robustness-efficiency trade-offs and exhibit other appealing 
properties, such as their direct applicability to {\em any\/} 
log-location-scale distribution and low computational costs. 
Specific contributions of this paper to the advancement of 
the QLS methodology include:
\begin{enumerate}
   \item An introduction of a nonlinear quantiles-based regression 
framework, which is valid for general loss models, and development 
of the log-linear approximation of its solution for the parameters 
of log-location-scale distributions.

  \item An extension of the aforementioned approximation for Weibull 
and Pareto $I$ models, for which the logarithmic transformation of 
quantiles does not isolate the parameters of interest. Thus, deriving 
log-oQLS/log-gQLS and their properties requires further work (see 
Section 2.4).

  \item An exploration of quantile-selection strategies, and their 
effect on the estimators' accuracy.

  \item A proposal of a new model-based outlier-screening rule.
Its effectiveness is studied via simulated and real data examples, 
and compared with that of some existing rules.
\end{enumerate}

The rest of the paper is organized as follows. 
In Section 2, we introduce nonlinear and linear regression models 
based on sample and theoretical quantiles and log-quantiles. 
Then, we define the oQLS/gQLS and log-oQLS/log-gQLS estimators for 
log-location-scale distributions, and compare them via simulations.
Further, we investigate robustness and efficiency properties of 
the log-QLS estimators, and search for optimal quantile-selection 
designs. We complete the section by constructing a goodness-of-fit 
test for model validation.
In Section 3, simulation studies are conducted to verify the 
log-oQLS and log-gQLS estimators' performance under data 
contamination and their effectiveness when applied to several 
outlier-screening rules. The power properties of the goodness-of-fit 
test are also studied in this section. 
Section 4 examines the practical performance of the new estimators 
and the goodness-of-fit test using real data. A summary of the paper 
and concluding remarks are provided in Section 5.

\section{Quantile Least Squares}

In this section, a general formulation of the {\em quantile least 
squares\/} (QLS) estimators for parametric distributions with 
the support on $(0, \infty)$ is presented. Two approaches are 
considered: the QLS based on raw quantiles, which results in nonlinear 
regression, and the log-QLS based on logarithmically transformed quantiles, 
which results in linear regression. Under both regression frameworks, two 
types of estimators -- ordinary and generalized least squares -- are 
studied for the class of log-location-scale families. Then, the QLS and 
log-QLS are compared using several statistical and computational metrics. 
Further, asymptotic robustness and efficiency properties of the log-QLS 
estimators are established and illustrated. Finally, a goodness-of-fit 
test is constructed to validate the models fitted by log-gQLS.

Suppose a sample of {\em independent and identically distributed\/} 
({\em i.i.d.\/}) positive continuous random variables, 
$X_1, \ldots, X_n$, is observed. 
Let the cumulative distribution function (cdf) 
$F_{\mbox{\boldmath\scriptsize $\theta$}}$, 
probability density function (pdf) 
$f_{\mbox{\boldmath\scriptsize $\theta$}}$, and 
the quantile function (qf) 
$F^{-1}_{\mbox{\boldmath\scriptsize $\theta$}}$ of these variables 
be given in a parametric form, with the (column) vector parameter
$\mbox{\boldmath $\theta$} = (\theta_1, \ldots, \theta_m)'$.
Further, let $X_{(1)} \leq \cdots \leq X_{(n)}$ denote the ordered 
sample values. The empirical estimator of the $p$th population 
quantile is the corresponding sample quantile
$X_{(\lceil n p \rceil)} = \widehat{F}^{-1}(p)$, where 
$\lceil \cdot \rceil$ denotes the rounding up operation. Also, 
throughout the paper the notation ${\cal AN}$ stands for 
``asymptotically normal.''

\subsection{QLS versus Log-QLS}

For log-location-scale loss models (Section 2.1.1), the QLS approach 
uses raw quantiles, which leads to nonlinear regression (Section 2.1.2). 
If quantiles are first logarithmically transformed, then the resulting 
model is linear regression (Section 2.1.3). Note that these regression 
frameworks are valid asymptotically, typically for sample sizes 
$n \geq 100$. The two approaches are compared in Section 2.1.4. 

\subsubsection{Log-Location-Scale Families}

For log-location-scale distributions, $m=2$ with
$\theta_1 = \mu$ ($-\infty < \mu < \infty$ is log-location) and 
$\theta_2 = \sigma$ ($\sigma > 0$ is log-scale). The pdf, cdf, and 
qf of these families are respectively given by:
\begin{equation}
f_{\mu, \sigma}(x) = 
\frac{1}{x \sigma} f_* \left( \frac{\log (x) - \mu}{\sigma} \right), 
\qquad
F_{\mu, \sigma}(x) = 
F_* \left( \frac{\log (x) -\mu}{\sigma} \right),
\qquad
F^{-1}_{\mu, \sigma}(u) = 
e^{ \mu + \sigma F_*^{-1}(u) },
\label{func}
\end{equation}
where $x > 0$ and $0 < u < 1$.
The functions $f_*$, $F_*$, $F_*^{-1}$ represent pdf, cdf, qf, 
respectively, of the corresponding standard (with $\mu=0$ and $\sigma=1$) 
location-scale family. By choosing $\mu$ or $\sigma$ known, the 
log-location-scale family simplifies to either the {\em log-scale\/} 
or {\em log-location\/} family, respectively. 

In Table 2.1, we list key facts for several log-location-scale families. 
In particular, the pdf $f_*$ and qf $F_*^{-1}$ are provided and, to 
facilitate comparisons with the maximum likelihood estimators (MLE), 
a standard version of the information matrix $\mathbf{I_*}$ is specified. 
All the selected distributions exhibit log-symmetric shapes (i.e., after
the underlying variable is log-transformed its pdf becomes symmetric).
Other popular choices of severity models such as Pareto $I$ and Weibull 
are log-skewed and require some modifications of the approach described 
in Section 2.1. They will be studied in Section 2.4.

\begin{center}
{\sc Table 2.1.} Key probabilistic formulas and information 
for selected log-location-scale families.

\medskip

\begin{tabular}{|c|c|c|c|}
\hline
Probability & Standard {\sc pdf} & Standard {\sc qf} & 
Information Matrix \\[-0.5ex]
Distribution & $f_*(z)$ & $F_*^{-1}(u)$ & 
$\mathbf{I_*} \, ( = \sigma^2 \times \mathbf{I} )$ \\
\hline
\hline
Log-Cauchy & $\dfrac{1}{\pi (1 + z^2)}$ & 
$\tan (\pi (u-0.5))$ & 
$\begin{bmatrix}
\frac{1}{2} & 0 \\
0 & \frac{1}{2} \\
\end{bmatrix}$ \\
Log-Laplace & $0.5 \, e^{-|z|}$ & 
$\left\{ 
\begin{array}{cl} 
 \log (2 u), & u \leq 0.5
\\[0.25ex]
 -\log (2 (1-u)), & u > 0.5
\end{array}
\right.
$
& 
$\begin{bmatrix}
1 & 0 \\
0 & 1 \\
\end{bmatrix}$ \\
Log-Logistic & $\dfrac{e^{-z}}{(1+e^{-z})^2}$ & 
$-\log (1/u-1)$ & 
$\begin{bmatrix}
\frac{1}{3} & 0 \\
0 & \frac{3+\pi^2}{9} \\
\end{bmatrix}$ \\[1ex]
Lognormal & $\frac{1}{\sqrt{2 \pi}} \, e^{-z^2/2}$ & 
$\Phi^{-1}(u)$ & 
$\begin{bmatrix}
1 & 0 \\
0 & 2 \\
\end{bmatrix}$ \\
\hline
\end{tabular}
\end{center}

\medskip

\subsubsection{Nonlinear Regression}

For a sample of size $n$ of continuous {\em  i.i.d\/}. random 
variables, sample quantiles 
$\widehat{F}^{-1}(p_1), \ldots, \widehat{F}^{-1}(p_k)$
with $0 < p_1 < \cdots < p_k < 1$ are jointly asymptotically 
normal \citep[see][Theorem B, p. 80]{s02a}. Specifically, 
\begin{equation}
\mathbf{\widehat{Q}} :=~ 
\big( \widehat{F}^{-1}(p_1), \ldots, \widehat{F}^{-1}(p_k) \big)'
~ \mbox{ is } ~ {\cal AN} 
\Big(
\mbox{\boldmath $\mu_{\theta}$}, \, n^{-1} \mbox{\boldmath $\Sigma_{\theta}$}
\Big) \, ,
\label{quant}
\end{equation}
where $\mbox{\boldmath $\mu_{\theta}$} = 
\big( F_{\mbox{\boldmath\scriptsize $\theta$}}^{-1}(p_1), \ldots, 
F_{\mbox{\boldmath\scriptsize $\theta$}}^{-1}(p_k) \big)'$ and 
the entries of $\mbox{\boldmath $\Sigma_{\theta}$}$ are given by 
\begin{equation}
\sigma_{ij} ~=~ \frac{\min\{p_i, p_j\} - p_i p_j}
{f_{\mbox{\boldmath\scriptsize $\theta$}}
(F_{\mbox{\boldmath\scriptsize $\theta$}}^{-1}(p_i)) 
f_{\mbox{\boldmath\scriptsize $\theta$}}
(F_{\mbox{\boldmath\scriptsize $\theta$}}^{-1}(p_j))} ,
\qquad
i, j = 1, \ldots, k.
\label{emp-sigma}
\end{equation}
Therefore, for large samples, the statement \eqref{quant} can be 
interpreted as a nonlinear regression model with approximately 
normal but {\em correlated\/} error terms. That is,
\begin{equation}
\widehat{F}^{-1}(p_i) ~=~ 
F_{\mbox{\boldmath\scriptsize $\theta$}}^{-1}(p_i) + \varepsilon_i,
\qquad
i = 1, \ldots, k,
\label{nl-reg0}
\end{equation}
where $\mbox{\boldmath $\varepsilon$} = 
\left( \varepsilon_1, \ldots, \varepsilon_k \right)'$ is 
${\cal AN} \big( \mbox{\bf 0}, \, 
\mbox{\boldmath $\Sigma_{\theta}$} / n \big)$. 
Using the matrix notation, the weighted nonlinear least squares 
solution of \eqref{nl-reg0} is
\begin{equation}
\widehat{\mbox{\boldmath $\theta$}} ~=~
\operatorname*{arg\,min}_{\theta_1,\ldots,\theta_m}
\left\{
\left( \mathbf{\widehat{Q}} - \mbox{\boldmath $\mu_{\theta}$} \right)' 
\mathbf{W}
\left( \mathbf{\widehat{Q}} - \mbox{\boldmath $\mu_{\theta}$} \right)
\right\} ,
\label{nl-reg1}
\end{equation}
where $\mathbf{W}$ denotes a $k \times k$ matrix of weights that has 
to be selected by the researcher. Typical choices of $\mathbf{W}$ 
include $\mathbf{I}_k$ (this is a $k \times k$ identity matrix) and 
$\mbox{\boldmath $\Sigma^{-1}_{\theta}$}$, 
which respectively yield the ordinary (oQLS) and generalized (gQLS) 
least squares solution of $\mbox{\boldmath $\theta$}$. 
Note that since $\mbox{\boldmath $\theta$} = (\theta_1, \ldots, \theta_m)'$
is $m$-dimensional, the number of selected quantiles $k$ should satisfy 
$k \geq m$. 

While at first glance this might look like a straightforward exercise, 
solving \eqref{nl-reg1} for general multi-parameter (e.g., $m \geq 3$) 
distributions is a non-trivial computational task, with various potential 
pitfalls such as numerical instability of optimization algorithms, 
multiple local minima, or poor choice of initial values 
\citep[see][Chapter 3]{bw88}. 
The iteratively reweighted least squares (IRLS) algorithm \citep[][]{rw94} 
is a standard numerical procedure for nonlinear regression. It relies on 
local linearization (i.e., a first-order Taylor approximation) of the model 
mean $\mbox{\boldmath $\mu_{\theta}$}$ around the true parameter value 
$\mbox{\boldmath ${\theta}$}^*$:
\begin{equation}
\mbox{\boldmath $\mu_{\theta}$} ~\approx~ \mbox{\boldmath $\mu_{\theta^*}$}
+ \mathbf{J_{\mbox{\boldmath\scriptsize $\! \theta^*$}}}
(\mbox{\boldmath $\theta$} -\mbox{\boldmath $\theta^*$}),
\label{taylor}
\end{equation}
where $\mathbf{J_{\mbox{\boldmath\scriptsize $\! \theta^*$}}}$ is a
$k \times m$ Jacobian matrix evaluated at $\mbox{\boldmath ${\theta}$}^*$.
The IRLS algorithm solves \eqref{nl-reg1} with 
$\mbox{\boldmath $\mu_{\theta}$}$ replaced by \eqref{taylor}, which 
results in a weighted least squares solution at each iteration. 
That is, starting with an initial guess 
$\mbox{\boldmath ${\theta}$}^{(0)}$,
the algorithm iterates until convergence (within a tolerance limit) 
according to 
\begin{equation}
\mbox{\boldmath ${\theta}$}^{(j+1)} =~ \mbox{\boldmath ${\theta}$}^{(j)}
+ (\mathbf{J'}_{\mbox{\boldmath\scriptsize $\! \! \theta$}^{(j)}} 
\mathbf{W} \,
\mathbf{J}_{\mbox{\boldmath\scriptsize $\! \theta$}^{(j)}})^{-1} 
\mathbf{J'}_{\mbox{\boldmath\scriptsize $\! \! \theta$}^{(j)}}
\mathbf{W} \,
(\mathbf{\widehat{Q}} - \mbox{\boldmath $\mu$}_{\boldsymbol{\theta}^{(j)}}), 
\qquad
j = 0, 1, 2, \ldots,
\label{irls}
\end{equation}
where typically $\mathbf{W} = \mathbf{I}_k$ or $\mathbf{W} = 
\mbox{\boldmath $\Sigma^{-1}$}_{\boldsymbol{\! \hspace{-4mm} \theta}^{(j)}}$,
yielding the oQLS and gQLS estimators of $\boldsymbol{\theta}$, respectively.

For the log-location-scale family parameterized by \eqref{func}, 
one could start, for example, with 
\[
\widehat{\sigma}^{(0)} = 
\frac{\log \widehat{F}^{-1}(0.75) - \log \widehat{F}^{-1}(0.25)}
{F_*^{-1}(0.75) - F_*^{-1}(0.25)},
\qquad
\widehat{\mu}^{(0)} = \log \widehat{F}^{-1}(0.75) - 
\widehat{\sigma}^{(0)} F_*^{-1}(0.75)
\]
and apply \eqref{irls} with the mean vector
\[
\boldsymbol{\mu}_{\mu^{(j)} \! , \, \sigma^{(j)}} = 
\left( F_{\mu^{(j)} \! , \, \sigma^{(j)}}^{-1}(p_1), \ldots,
F_{\mu^{(j)} \! , \, \sigma^{(j)}}^{-1}(p_k) \right)'
\]
and the Jacobian matrix
\[
\mathbf{J}_{\mu^{(j)} \!, \, \sigma^{(j)}} ~=~ 
\begin{bmatrix}
\frac{\partial F_{\mu, \sigma}^{-1}(p_1)}{\partial \mu} & 
\frac{\partial F_{\mu, \sigma}^{-1}(p_1)}{\partial \sigma} \\[1ex]
 \vdots & \vdots \\[1ex]
\frac{\partial F_{\mu, \sigma}^{-1}(p_k)}{\partial \mu} & 
\frac{\partial F_{\mu, \sigma}^{-1}(p_k)}{\partial \sigma} \\
\end{bmatrix}_{\mu = \mu^{(j)} \! , \, \sigma = \sigma^{(j)}} =~
\begin{bmatrix}
F_{\mu^{(j)} \! , \, \sigma^{(j)}}^{-1}(p_1) & 
F_{\mu^{(j)} \! , \, \sigma^{(j)}}^{-1}(p_1) \, F_*^{-1}(p_1) \\[1ex]
 \vdots & \vdots \\[1ex]
F_{\mu^{(j)} \! , \, \sigma^{(j)}}^{-1}(p_k) & 
F_{\mu^{(j)} \! , \, \sigma^{(j)}}^{-1}(p_k) \, F_*^{-1}(p_k) \\
\end{bmatrix}.
\]

\medskip

\noindent
{\bf Note 2.1} ~ {\em Numerical stabilization of IRLS\/}.
When the columns of $\mathbf{J}_{\boldsymbol{\theta}^{(j)}}$ 
are nearly collinear,
$\mathbf{J}' \mathbf{W} \mathbf{J}$ in \eqref{irls} is close 
to singular and the increment can grow uncontrollably, pushing 
the iterates into implausible regions of the parameter space 
\citep[Section 3.5.2]{bw88}. A standard safeguard is the 
Levenberg--Marquardt compromise \citep[see][]{l44,m63},
which stabilizes the step by adding a damping term to the 
cross-product matrix:
\[
\boldsymbol{\theta}^{(j+1)} =~ \boldsymbol{\theta}^{(j)}
+ \big( \mathbf{J}'_{\boldsymbol{\theta}^{(j)}} \mathbf{W} \,
\mathbf{J}_{\boldsymbol{\theta}^{(j)}} + \lambda \,
\mathbf{D}_{\boldsymbol{\theta}^{(j)}} \big)^{-1}
\mathbf{J}'_{\boldsymbol{\theta}^{(j)}} \mathbf{W} \,
(\mathbf{\widehat{Q}} - \boldsymbol{\mu}_{\boldsymbol{\theta}^{(j)}}),
\]
where $\mathbf{D}_{\boldsymbol{\theta}^{(j)}} = \operatorname{diag}
(\mathbf{J}'_{\boldsymbol{\theta}^{(j)}} \mathbf{W}
\mathbf{J}_{\boldsymbol{\theta}^{(j)}})$ and the conditioning factor
$\lambda \geq 0$ is raised until the step decreases \eqref{nl-reg1} 
and lowered otherwise; $\lambda \to 0$ recovers the 
Gauss--Newton step in \eqref{irls}, while $\lambda \to \infty$ 
tends to a scaled steepest-descent direction.
\hfill $\Box$

\medskip

For parametric distributions with complicated quantile function
expressions, the derivatives of qf with respect to the parameters may 
be difficult to work with. In those cases, direct search algorithms 
such as the Nelder-Mead (NM) method may be preferred. For example, 
\citet[][]{xic14} used the NM method for computing the oQLS estimators 
of the $g$-and-$h$ family parameters. Both algorithms -- IRLS and NM -- 
will be implemented for the log-location-scale distributions and 
compared in Section 2.1.4.

\subsubsection{Linear Regression}

To improve convergence of the iterative algorithms and overall 
performance of nonlinear regression analysis, it is a common practice 
to linearize the model (if possible) by first transforming the response 
variable or the model parameters or both. For the log-location-scale 
family, the logarithmic transformation of the response variable (i.e., 
the selected quantiles) is especially effective.

As specified in \eqref{func}, the support of $f_{\mu, \sigma}$ 
is $(0, \infty)$ and thus $\widehat{F}^{-1}(p) > 0$ and 
$F_{\mu, \sigma}^{-1}(p) > 0$ for $0 < p < 1$. Therefore, 
the transformations $\log \widehat{F}^{-1}(p)$ and 
$\log F_{\mu, \sigma}^{-1}(p)$ are well defined. An application 
of the delta method \citep[see][Section 3.3]{s02a} 
to the log-transformed \eqref{quant}, with 
$\boldsymbol{\theta} = (\mu, \sigma)$, implies that 
\begin{equation}
\log \mathbf{\widehat{Q}} ~=~
\big( \log \widehat{F}^{-1}(p_1), \ldots, \log \widehat{F}^{-1}(p_k) \big)'
~ \mbox{ is } ~ {\cal AN} 
\Big(
\mbox{\boldmath $\mu$}_{\mu, \sigma}, \, n^{-1}
\mbox{\boldmath $D$}_{\mu, \sigma} \mbox{\boldmath $\Sigma$}_{\mu, \sigma} 
\mbox{\boldmath $D'$}_{\! \! \mu, \sigma}
\Big) \, ,
\label{log-quant}
\end{equation}
where $\mbox{\boldmath $\mu$}_{\mu, \sigma} = \big( 
\log F_{\mu, \sigma}^{-1}(p_1), \ldots, \log F_{\mu, \sigma}^{-1}(p_k) \big)' = 
\big( \mu + \sigma F_*^{-1}(p_1), \ldots,  \mu + \sigma F_*^{-1}(p_k) \big)'$,
the Jacobian $\mbox{\boldmath $D$}_{\mu, \sigma}$ is a diagonal matrix with 
elements $1/F_{\mu, \sigma}^{-1}(p_i)$ and the entries of 
$\mbox{\boldmath $\Sigma$}_{\mu, \sigma}$ are defined by \eqref{emp-sigma}.
Moreover, 
$\mbox{\boldmath $D$}_{\mu, \sigma} \mbox{\boldmath $\Sigma$}_{\mu, \sigma} 
\mbox{\boldmath $D'$}_{\! \! \mu, \sigma} = \sigma^2 \boldsymbol{\Sigma_*}$, 
where the entries of $\boldsymbol{\Sigma_*}$ are known and given by
\begin{equation}
\sigma_{ij}^* ~=~ 
\frac{\min\{ p_i, p_j \} - p_i p_j} 
{f_*(F_*^{-1}(p_i)) f_*(F_*^{-1}(p_j))} ,
\qquad
i, j = 1, \ldots, k.
\label{Sigma*}
\end{equation}

Now, based on \eqref{log-quant}, the approximation \eqref{taylor} for 
the weighted least squares solution \eqref{nl-reg1} is {\em exact\/}
and thus \eqref{irls} converges in one step. More specifically, since 
the Jacobian matrix is constant,
\begin{equation}
\mathbf{J}_{\mu^{(j)} \!, \, \sigma^{(j)}} ~=~ 
\begin{bmatrix}
\frac{\partial \log F_{\mu, \sigma}^{-1}(p_1)}{\partial \mu} & 
\frac{\partial \log F_{\mu, \sigma}^{-1}(p_1)}{\partial \sigma} \\[1ex]
 \vdots & \vdots \\[1ex]
\frac{\partial \log F_{\mu, \sigma}^{-1}(p_k)}{\partial \mu} & 
\frac{\partial \log F_{\mu, \sigma}^{-1}(p_k)}{\partial \sigma} \\
\end{bmatrix}_{\mu = \mu^{(j)} \! , \, \sigma = \sigma^{(j)}} =~
\begin{bmatrix}
1 & F_*^{-1}(p_1) \\[1ex]
 \vdots & \vdots \\[1ex]
1 & F_*^{-1}(p_k) \\
\end{bmatrix}
~=: \mathbf{X},
\label{matrixX}
\end{equation}
the iteration \eqref{irls} simplifies to
$
\mbox{\boldmath ${\widehat{\beta}}$} ~=~ 
(\mathbf{X'} \mathbf{W} \, \mathbf{X})^{-1} 
\mathbf{X'} \mathbf{W} \, \mathbf{Y} ,
$
where $\mbox{\boldmath ${\widehat{\beta}}$} = 
(\widehat{\mu}, \widehat{\sigma})' =
\mbox{\boldmath ${\theta}$}^{(j)}$ for $j = 1, 2, 3, \ldots$ 
and
$\mathbf{Y} = \mathbf{\log \widehat{Q}}$.
Choosing $\mathbf{W} = \mathbf{I}_k$ or $\mathbf{W} = 
\big( \mbox{\boldmath $D$}_{\mu, \sigma} 
\mbox{\boldmath $\Sigma$}_{\mu, \sigma} 
\mbox{\boldmath $D'$}_{\! \! \mu, \sigma} \big)^{-1} = 
\sigma^{-2} \boldsymbol{\Sigma_*^{-1}}$,
respectively yields
\begin{equation}
\widehat{\mbox{\boldmath $\beta$}}_{\mbox{\tiny log-oQLS}} ~=~
(\mathbf{X'} \mathbf{X})^{-1} \mathbf{X'} \mathbf{Y} 
\label{oQLS}
\end{equation}
and
\begin{equation}
\widehat{\mbox{\boldmath $\beta$}}_{\mbox{\tiny log-gQLS}} ~=~
(\mathbf{X'} \mbox{\boldmath $\Sigma_*^{-1}$} \mathbf{X})^{-1} 
\mathbf{X'}  \mbox{\boldmath $\Sigma_*^{-1}$} \mathbf{Y} .
\label{gQLS}
\end{equation}
Note that \eqref{oQLS} and \eqref{gQLS} represent the ordinary 
and generalized least squares estimators of 
$\boldsymbol{\beta} = (\mu, \sigma)'$ in the {\em linear\/} 
regression model
\begin{equation}
\mathbf{Y} ~=~ \mathbf{X} \mbox{\boldmath $\beta$} + 
\mbox{\boldmath $\varepsilon$},
\label{l-reg}
\end{equation}
where $\mathbf{Y} = \mathbf{\log \widehat{Q}}$ denotes data
(response variable), the design matrix $\mathbf{X}$ is given 
by \eqref{matrixX},
the error term $\mbox{\boldmath $\varepsilon$} = 
\left( \varepsilon_1, \ldots, \varepsilon_k \right)'$ is 
${\cal AN} \big( \mbox{\bf 0}, \, 
\sigma^2 \boldsymbol{\Sigma_*} / n \big)$,
and $\boldsymbol{\Sigma_*}$ is specified by \eqref{Sigma*}.

\subsubsection{Numerical Comparisons}

In this section, the nonlinear (QLS) and linear (log-QLS) regression
estimators of $\mu$ and $\sigma$ are compared using a small-scale 
simulation study. The study design is based on the following choices:
\begin{itemize}
  \item {\em Data-generating distributions\/}. All the log-location-scale
distributions listed in Table 2.1 with $\mu =0$ and $\sigma =1$.

\vspace{-1ex}

  \item {\em Estimators\/}. oQLS, gQLS, log-oQLS, log-gQLS, based on
$k=25$ quantiles with the probability levels $p_i$ uniformly spaced 
between 0.05 and 0.95: $p_i = 0.05 + 0.90 (i-1) / 24$, $i = 1, \ldots, 25$.
The nonlinear estimators oQLS and gQLS are computed using the IRLS and NM 
algorithms.

\vspace{-1ex}

  \item {\em Sample sizes\/}. ~$n = 10^2, \, 10^3$.
  
\vspace{-1ex}
  
  \item {\em Number of Monte Carlo runs\/}. ~$M = 10^4$.
\end{itemize}

Table 2.2 summarizes the results of the simulation study. It is clear 
from the table that the NM algorithm ($\widehat{\mu}_2, \widehat{\sigma}_2$) 
is the weakest among the three approaches in terms of bias and root-MSE. 
For some distributions, computations of the IRLS algorithm 
($\widehat{\mu}_1, \widehat{\sigma}_1$) required the Levenberg--Marquardt 
conditioning factor (we started with $\lambda = 10^{-2}$ and made changes 
by a multiplicative factor of 10 in subsequent steps; see Note 2.1), but 
overall the method performed well. Nonetheless, it was further improved 
by the corresponding log-QLS estimators ($\widehat{\mu}_3, \widehat{\sigma}_3$)
which showed least variability, almost no bias, and worked especially well 
for heavy-tailed log-Cauchy.

\newpage

\begin{center}
{\sc Table 2.2.} Absolute bias and root-MSE of $\widehat{\mu}$ and 
$\widehat{\sigma}$ for selected log-location-scale distributions
\\[-1ex]
and sample sizes $n$. The estimators are computed using
IRLS $(\widehat{\mu}_1, \widehat{\sigma}_1)$, 
NM $(\widehat{\mu}_2, \widehat{\sigma}_2)$,
\\[-1ex]
and log-QLS $(\widehat{\mu}_3, \widehat{\sigma}_3)$.
True parameter values are $\mu = 0$ and $\sigma = 1$.

\medskip

\begin{tabular}{|c|c|ccc|ccc|ccc|ccc|}
\hline
Probability & $n$ & 
\multicolumn{6}{|c|}{Absolute Bias} &
\multicolumn{6}{|c|}{Root-MSE} \\[-0.5ex]
\cline{3-14}
Distribution & & $\widehat{\mu}_1$ & $\widehat{\mu}_2$ & $\widehat{\mu}_3$ &
 $\widehat{\sigma}_1$ & $\widehat{\sigma}_2$ & $\widehat{\sigma}_3$ &
 $\widehat{\mu}_1$ & $\widehat{\mu}_2$ & $\widehat{\mu}_3$ &
 $\widehat{\sigma}_1$ & $\widehat{\sigma}_2$ & $\widehat{\sigma}_3$ \\
\hline
\multicolumn{14}{l}{oQLS and log-oQLS} \\
\hline
{\em Log-Cauchy} & $10^2$ & 0.29 & 0.32 & 0.07 & 0.05 & 0.05 & 0.10 & 
 2.12 & 1.97 & 0.43 & 0.71 & 0.68 & 0.33 \\
 & $10^3$ & 0.01 & 0.02 & 0.01 & 0.00 & 0.00 & 0.01 &
 0.62 & 0.63 & 0.11 & 0.22 & 0.22 & 0.08 \\
{\em Log-Laplace} & $10^2$ & 0.02 & 0.02 & 0.01 & 0.03 & 0.03 & 0.01 &
 0.24 & 0.24 & 0.13 & 0.22 & 0.22 & 0.11 \\
 & $10^3$ & 0.00 & 0.00 & 0.00 & 0.00 & 0.00 & 0.00 &
 0.06 & 0.06 & 0.04 & 0.07 & 0.07 & 0.03 \\
{\em Log-Logistic} & $10^2$ & 0.05 & 0.05 & 0.01 & 0.04 & 0.04 & 0.00 &
 0.41 & 0.41 & 0.18 & 0.23 & 0.23 & 0.09 \\
 & $10^3$ & 0.00 & 0.00 & 0.00 & 0.00 & 0.00 & 0.00 & 
 0.12 & 0.12 & 0.05 & 0.07 & 0.07 & 0.03 \\
{\em Lognormal} & $10^2$ & 0.00 & 0.00 & 0.01 & 0.02 & 0.02 & 0.00 & 
 0.13 & 0.13 & 0.10 & 0.13 & 0.13 & 0.08 \\
 & $10^3$ & 0.00 & 0.00 & 0.00 & 0.00 & 0.00 & 0.00 & 0.04 & 0.04 & 
 0.03 & 0.04 & 0.04 & 0.03 \\
\hline
\multicolumn{14}{l}{gQLS and log-gQLS} \\
\hline
{\em Log-Cauchy} & $10^2$ & 0.11 & 2.71 & 0.00 & 0.07 & ** & 0.03 &
 0.63 & 9.69 & 0.15 & 0.24 & ** & 0.15 \\
 & $10^3$ & 0.01 & 0.06 & 0.00 & 0.01 & 0.01 & 0.00 & 
 0.05 & 1.35 & 0.04 & 0.05 & 0.19 & 0.04 \\
{\em Log-Laplace} & $10^2$ & 0.01 & 0.02 & 0.01 & 0.01 & 0.01 & 0.01 &
 0.11 & 0.28 & 0.11 & 0.11 & 0.16 & 0.11 \\
 & $10^3$ & 0.00 & 0.00 & 0.00 & 0.00 & 0.00 & 0.00 & 
 0.03 & 0.03 & 0.03 & 0.03 & 0.03 & 0.03 \\
{\em Log-Logistic} & $10^2$ & 0.00 & 0.04 & 0.01 & 0.00 & 0.01 & 0.00 &
 0.18 & 1.56 & 0.18 & 0.09 & 0.54 & 0.09 \\
 & $10^3$ & 0.00 & 0.00 & 0.00 & 0.00 & 0.00 & 0.00 &
 0.05 & 0.05 & 0.05 & 0.03 & 0.03 & 0.03 \\
{\em Lognormal} & $10^2$ & 0.01 & 0.01 & 0.01 & 0.00 & 0.00 & 0.00 &
 0.10 & 0.10 & 0.10 & 0.08 & 0.08 & 0.08 \\
 & $10^3$ & 0.00 & 0.00 & 0.00 & 0.00 & 0.00 & 0.00 &
 0.03 & 0.03 & 0.03 & 0.03 & 0.03 & 0.03 \\
\hline
\multicolumn{14}{l}{\footnotesize $**$ ~For estimation of log-Cauchy 
$\sigma$ when $n=100$, the NM algorithm failed to converge in 
2,000 iterations.} \\
\end{tabular}
\end{center}

\medskip

In addition, in Table 2.3 we report computational times for gQLS 
and log-gQLS estimators, when sample size $n$ grows from $10^6$ 
to $10^9$. (The corresponding oQLS and log-oQLS estimators take 
similar times to compute as their gQLS counterparts.) Following 
\citet[][]{ab25}, we performed all computations using
$\mbox{MATLAB}^{\copyright}$ R2025b software but on a faster 
laptop (with Apple M3 8-core CPU, RAM 16GB, and Mac OS). 
The table demonstrates that the NM method is the slowest among 
the three, IRLS is significantly faster, and log-QLS is the fastest. 
Interestingly, the regression framework proposed here even speeds 
up the numerical algorithms because their iterations are done on 
$k=25$ points as opposed to $n$ which would be required by 
non-explicit MLEs. 
For instance, for $n = 10^9$ at log-Logistic the IRLS and NM 
procedures were computed in 237 and 457 seconds, respectively, 
and at log-Cauchy, the respective computing times were 133 and 
469. But as reported by \citet[][Table 3]{ab25}, at Logistic it 
took 28,461 seconds to compute MLE, and at Cauchy the estimator 
failed to converge over a span of several days. 

\begin{center}
{\sc Table 2.3.} Computational times (in seconds) of 
gQLS and log-gQLS estimators for large $n$.

\medskip

\begin{tabular}{|c|c|cccc|}
\hline
Sample & Estimation & 
\multicolumn{4}{|c|}{Probability Distribution} \\[-0.5ex]
Size & Method & 
Log-Cauchy & Log-Laplace & Log-Logistic & Lognormal \\
\hline
\hline
$n = 10^6$ & $\mbox{gQLS}_{\mbox{\tiny IRLS}}$ & 0.05 & 0.08 & 0.08 & 0.07 \\
 & $\mbox{gQLS}_{\mbox{\tiny NM}}$ & 0.31 & 0.44 & 0.44 & 0.32 \\
 & log-gQLS & {\bf 0.03} & {\bf 0.06} & {\bf 0.06} & {\bf 0.05} \\
\hline
$n = 10^7$ & $\mbox{gQLS}_{\mbox{\tiny IRLS}}$ & 0.28 & 0.49 & 0.31 & 0.39 \\
 & $\mbox{gQLS}_{\mbox{\tiny NM}}$ & 0.75 & 0.90 & 0.70 & 0.64 \\
 & log-gQLS & {\bf 0.22} & {\bf 0.35} & {\bf 0.25} & {\bf 0.33}\\
\hline
$n = 10^8$ & $\mbox{gQLS}_{\mbox{\tiny IRLS}}$ & 2.58 & 5.12 & 2.91 & 3.74 \\
 & $\mbox{gQLS}_{\mbox{\tiny NM}}$ & 9.63 & 8.30 & 5.81 & 3.99 \\
 & log-gQLS & {\bf 2.29} & {\bf 4.00} & {\bf 2.34} & {\bf 3.38} \\
\hline
$n = 10^9$ & $\mbox{gQLS}_{\mbox{\tiny IRLS}}$ & 133 & 194 & 237 & 223 \\
 & $\mbox{gQLS}_{\mbox{\tiny NM}}$ & 469 & 797 & 457 & 379 \\
 & log-gQLS & {\bf 118} & {\bf 178} & {\bf 99} & {\bf 199} \\
\hline
\end{tabular}
\end{center}

\medskip

\subsection{Properties of Log-QLS}

As is evident from Tables 2.2 and 2.3, for the log-location-scale family 
the log-QLS-type estimators are more accurate and computationally cheaper 
than their QLS counterparts. Therefore, for the remainder of the paper 
our focus will be on log-oQLS and log-gQLS.

\subsubsection{Asymptotic Normality}

It follows from derivations of \eqref{oQLS} and \eqref{gQLS} that the 
log-QLS estimators are the QLS estimators for location-scale distributions, 
only based on log-transformed quantiles. Therefore, their asymptotic 
distributions carry over from Sections 2 and 3 of \citet{ab25}, with 
some modifications due to the log-transformation of data. Specifically: 
\begin{equation}
\widehat{\mbox{\boldmath $\beta$}}_{\mbox{\tiny log-oQLS}}
~ \mbox{ is } ~ {\cal AN} 
\left(
\mbox{\boldmath $\beta$}, \, 
\dfrac{\sigma^2}{n} \, (\mathbf{X' X})^{-1} \mathbf{X'} 
\mbox{\boldmath $\Sigma_*$}
\mathbf{X} (\mathbf{X' X})^{-1}
\right)
\label{l-an-o}
\end{equation}
and
\begin{equation}
\widehat{\mbox{\boldmath $\beta$}}_{\mbox{\tiny log-gQLS}}
~ \mbox{ is } ~ {\cal AN} 
\left(
\mbox{\boldmath $\beta$}, \, 
\dfrac{\sigma^2}{n} \, 
\big( \mathbf{X'} \mbox{\boldmath $\Sigma_*^{-1}$} \mathbf{X} \big)^{-1}
\right) ,
\label{l-an-g}
\end{equation}
where $\widehat{\mbox{\boldmath $\beta$}}_{\mbox{\tiny log-oQLS}}$
and $\widehat{\mbox{\boldmath $\beta$}}_{\mbox{\tiny log-gQLS}}$
are defined by \eqref{oQLS} and \eqref{gQLS}, respectively,
and {\boldmath $\beta$} $= (\mu, \sigma)'$.

\medskip

\noindent
{\bf Note 2.2} ~ {\em One-parameter models\/}.
If a one-parameter family -- log-location or log-scale -- needs 
to be estimated, the statements \eqref{oQLS}--\eqref{l-an-g} still 
remain valid, but the design matrix \eqref{matrixX} will be a column 
of 1's (for log-location) or a column of $F_*^{-1}(p)$'s (for log-scale). 
Also, the data vector will be modified to include the known parameter. 
For example, to estimate $\sigma$ when $\mu$ is known, one should 
define 
$\mathbf{Y} = 
\left( \log \big( \widehat{F}^{-1}(p_1) \big) - \mu, \ldots, 
\, \log \big( \widehat{F}^{-1}(p_k) \big) - \mu \right)'$.
\hfill $\Box$

\subsubsection{Robustness}

For QLS and log-QLS estimators, it is clear from the choice of quantile 
probability levels,
\[
0 < a = p_1 < p_2 < \cdots < p_{k-1} < p_k = b < 1,
\]
that the order statistics with the index below $\lceil n a \rceil$ 
and above $\lceil n b \rceil$ play no explicit role in estimation 
of the regression model \eqref{nl-reg0}. This implies that QLS and 
log-QLS estimators are globally robust with the (asymptotic) breakdown 
point (BP) equal to:
\begin{equation}
\mbox{BP} ~=~ 
\min \left\{ \mbox{LBP}, \mbox{UBP} \right\} 
~=~ \min \left\{ a, 1-b \right\} > 0.
\label{nl-bp}
\end{equation}
Note that when the underlying probability distribution 
$F_{\mu, \sigma}$ is skewed, it is reasonable to consider 
{\em lower\/} (LBP) and {\em upper\/} (UBP) breakdown points 
separately. For more details on the relevance of LBP and UBP 
in insurance, see \citet[][]{bs00} and \citet[][]{s02b}.

The influence function (IF) measures the approximate numerical 
impact of the observation $X_i = x$ on the error of estimation 
of $\boldsymbol{\beta} = (\mu, \sigma)'$. Taking into account 
the logarithmic transformation of quantiles, the IFs of 
$\widehat{\mbox{\boldmath $\beta$}}_{\mbox{\tiny log-oQLS}}$ 
and 
$\widehat{\mbox{\boldmath $\beta$}}_{\mbox{\tiny log-gQLS}}$ 
are given by
\[
\mbox{IF} 
\big( 
x, \widehat{\mbox{\boldmath $\beta$}}_{\mbox{\tiny log-oQLS}} 
\big) ~=~
(\mathbf{X' X})^{-1} \mathbf{X'} 
\left( \mbox{IF} \big( x, \widehat{F}^{-1}(p_1) \big), \ldots,
\mbox{IF} \big( x, \widehat{F}^{-1}(p_k) \big) \right)'
\]
and
\[
\mbox{IF} 
\big( 
x, \widehat{\mbox{\boldmath $\beta$}}_{\mbox{\tiny log-gQLS}} 
\big) ~=~
(\mathbf{X'} \mbox{\boldmath $\Sigma_*^{-1}$} \mathbf{X})^{-1} 
\mathbf{X'} \mbox{\boldmath $\Sigma_*^{-1}$}  
\left( \mbox{IF} \big( x, \widehat{F}^{-1}(p_1) \big), \ldots,
\mbox{IF} \big( x, \widehat{F}^{-1}(p_k) \big) \right)',
\]
where $x > 0$,
$
\mbox{IF} \big( x, \widehat{F}^{-1}(p_i) \big) = 
\sigma F_{\mu, \sigma}^{-1}(p_i) \, \frac{p_i - \mbox{\bf 1} 
\{ x \, \leq \, F_{\mu, \sigma}^{-1}(p_i) \}}{f_*(F_*^{-1}(p_i))},
~ i = 1, \ldots, k,
$
and $\mbox{\bf\large 1} \{ \cdot \}$ denotes the indicator function.
Here, the condition \eqref{nl-bp} ensures that all 
$\mbox{IF} \big( x, \widehat{F}^{-1}(p_i) \big)$'s and 
the corresponding weight functions that multiply them are bounded.
This makes
$\widehat{\mbox{\boldmath $\beta$}}_{\mbox{\tiny log-oQLS}}$ 
and 
$\widehat{\mbox{\boldmath $\beta$}}_{\mbox{\tiny log-gQLS}}$ 
locally robust. 

In Figure 2.1, we plot 
$\mbox{IF} 
\big( x, \widehat{\mbox{\boldmath $\beta$}}_{\mbox{\tiny log-oQLS}} \big)$
and
$\mbox{IF} 
\big( x, \widehat{\mbox{\boldmath $\beta$}}_{\mbox{\tiny log-gQLS}} \big)$
for selected log-location-scale families using $\mu = 0$ and $\sigma = 1$.
As illustrations in \citet[][Chapter 2]{hrrs86} suggest, the shapes 
of these IFs look like stepwise approximations of a {\em half of the IFs 
of various robust estimators for location and scale parameters\/}.
But notice that there is some influence capping near $x = 0$, the point 
of infinite magnitude for logarithmically transformed data.
For estimation of $\sigma$, they behave like an approximate version of 
an $M$-estimator for scale \citep[][Figure 2, p.123]{hrrs86}, with 
log-gQLS exhibiting a slightly sharper dip around $x = 1$ (equivalently, 
$\mu = 0 = \log (1)$). 
For estimation of $\mu$, the log-oQLS estimators act like a trimmed/winsorized 
mean or Huber estimator \citep[][Figure 1, p.105]{hrrs86}, and the IF of 
the log-gQLS estimators exhibits a variety of shapes: for lognormal and 
log-Logistic distributions, it acts like a trimmed/winsorized mean; 
for log-Laplace, it behaves like a median; and for log-Cauchy, its shape 
resembles that of a Tukey's biweight 
\citep[][Figure 3, p.151]{hrrs86}. We note in passing that the latter is 
a prominent example of the class of {\em redescending\/} $M$-estimators 
that is known to be an effective outlier rejection technique. However, 
the main difference between log-QLS methodology and $M$-estimators is this: 
For $M$-estimators, one has to ``know'' and specify the right shape of
the IF, while in log-QLS estimation the IF shapes emerge naturally with 
no intervention from the researcher.

\begin{center}
\resizebox{165mm}{135mm}{\includegraphics{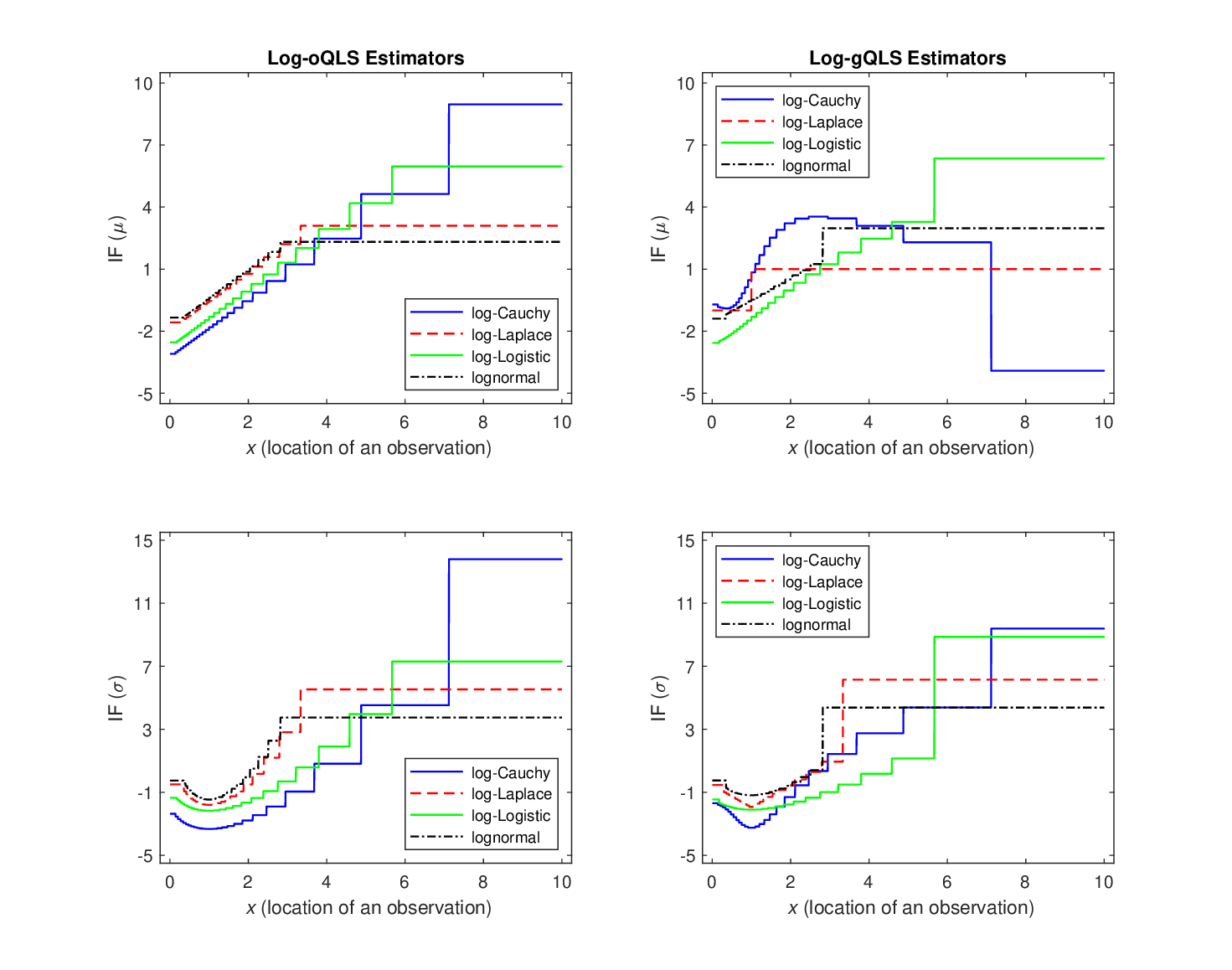}}
\\[-3ex]
{\sc Figure 2.1.} ~Influence functions of the log-oQLS and 
log-gQLS estimators of $\mu$ and $\sigma$ 
\\[-1ex]
for log-Cauchy, log-Laplace, log-Logistic, and lognormal 
distributions. The quantiles 
\\[-1ex]
are selected according to \eqref{quants} 
with $(a, b) = (0.15, 0.85)$ and $k = 25$.
\end{center}

\medskip

\noindent
{\bf Note 2.3} ~ {\em Interpretation of the IF sign\/}.
If the influence function is {\em negative\/}/{\em positive\/} 
at point $x$, then the sample observation located at $x$ 
{\em underestimates\/}/{\em overestimates\/} the corresponding 
parameter. For example, for all estimators in Figure 2.1, IFs 
are negative when $x$ is near $0$ leading to underestimation of 
$\mu = 0$ (because $\log x < 0$ when $0 < x < 1$), as well as 
to underestimation of $\sigma = 1$. Interestingly, for the log-gQLS 
estimator of $\mu$ under the log-Cauchy model, large values of 
$x$ also lead to underestimation of the parameter. Finally, note 
that the right-most step of each IF curve remains at the same 
level as $x \rightarrow \infty$ (i.e., there are no more jumps),
which implies that a moderately large observation (e.g., $x=10$) 
and a huge one (e.g., $x=10^{10}$) have the same effect on 
the estimator.
\hfill $\Box$

\medskip

\noindent
{\bf Note 2.4} ~ {\em Other roles of $a$ and $b$\/}.
In addition to robustness considerations, the quantile confidence levels 
$a$ and $b$ can be chosen to zoom in certain parts of the sample. This 
may be required by a modeling problem. For example, the choice $a=0.01$ 
and $b=0.99$ results in a log-QLS estimator that is based on almost 
the entire range of the sample. Such an estimator acts on data like MLE. 
In insurance applications, however, one is often interested in estimators 
that capture the upper tail well. If the tail is viewed as top 50\% of 
the distribution, then $a=0.50$ and $b=0.99$ is a reasonable choice. 
If the tail is interpreted as top 20\% of the distribution, then $a=0.80$ 
and $b=0.99$ will do the job.
\hfill $\Box$

\subsubsection{Relative Efficiency and Quantile Selection}

For the log-location-scale family of Section 2.1.1, the standard 
regularity conditions on $F_{\mu, \sigma}$ are met 
\citep[see, for example,][Section 4.2.2]{s02a}, and 
thus the MLE of $\mbox{\boldmath $\beta$} = (\mu, \sigma)'$ satisfies:
\[
\widehat{\mbox{\boldmath $\beta$}}_{\mbox{\tiny MLE}}
~ \mbox{ is } ~ {\cal AN} 
\left(
\mbox{\boldmath $\beta$}, \, 
\dfrac{\sigma^2}{n} \, \mathbf{I^{-1}_*}
\right) ,
\]
where $\mathbf{I_*}$ is the Fisher information matrix (see Table 2.1). 
Since MLE is the most efficient ${\cal AN}$ estimator, its performance 
can be used as a benchmark for the log-oQLS and log-gQLS estimators of 
$\mbox{\boldmath $\beta$}$. In particular, we will employ the ratio of 
the {\em generalized variances\/} raised to the power $1/m$, where
$m$ is the dimension of the parameter vector, as the {\em asymptotic 
relative efficiency\/} (ARE) criterion \citep[][Section 4.1]{s02a}. 
For the log-location-scale family, $m=2$ and thus we have:
\begin{eqnarray}
\mbox{ARE} \, \big( \mbox{log-oQLS}, \, \mbox{MLE} \big) & = &
\left( \frac{\mbox{det} \left[ \dfrac{\sigma^2}{n} \, \mathbf I_*^{-1} \right]}
{\mbox{det} 
\left[ \dfrac{\sigma^2}{n} \,
(\mathbf{X' X})^{-1} \mathbf{X'} 
\mbox{\boldmath $\Sigma_*$}
\mathbf{X} (\mathbf{X' X})^{-1}
\right]} 
\right)^{1/2}
\nonumber
\\[1.5ex]
 & = &
\left( \frac{\mbox{det} \left[ \mathbf I_*^{-1} \right]}
{\mbox{det} 
\big[
(\mathbf{X' X})^{-1} \mathbf{X'} 
\mbox{\boldmath $\Sigma_*$}
\mathbf{X} (\mathbf{X' X})^{-1}
\big]} 
\right)^{1/2}
\label{AREo}
\end{eqnarray}
and
\begin{eqnarray}
\mbox{ARE} \, \big( \mbox{log-gQLS}, \, \mbox{MLE} \big) & = &
\left( \frac{\mbox{det} \left[ \dfrac{\sigma^2}{n} \,
\mathbf I_*^{-1} \right]}
{\mbox{det} 
\left[ \dfrac{\sigma^2}{n} \,
( \mathbf{X'} \mbox{\boldmath $\Sigma^{-1}_*$} \mathbf{X} )^{-1}
\right]} 
\right)^{1/2} 
\nonumber
\\[1.5ex]
 & = &
\left( \frac{\mbox{det} \left[ \mathbf I_*^{-1} \right]}
{\mbox{det} 
\big[
( \mathbf{X'} \mbox{\boldmath $\Sigma^{-1}_*$} \mathbf{X} )^{-1}
\big]} 
\right)^{1/2} ,
\label{AREg}
\end{eqnarray}
where `det' stands for the determinant of a square matrix.

Formulas \eqref{AREo} and \eqref{AREg} match the ARE formulas for 
location-scale distributions, and they depend on $k$ and how 
the levels $p_1, \ldots, p_k$ are chosen.
For example, by choosing $k=2$, one could construct a percentile 
matching estimator \citep[][Section 13.1]{kpw12}, which is robust 
but not efficient. On the other hand, \citet[][Section 3.3]{ab25} 
fixed $p_1 = a > 0$ and $p_{25} = b < 1$, used the uniform 
spacing of $p_2, \ldots, p_{24}$ over $[a,b]$, and showed that 
the gQLS approach is more efficient than oQLS, while offering 
the same degrees of robustness. Here we will check if other 
quantile-spacing designs can further improve the gQLS estimators' 
efficiency. Thus, starting with fixed $0 < a = p_1 < p_k = b < 1$, 
we consider the following strategies:
\begin{enumerate}
  \item {\em Uniformly-spaced probability levels\/} 
$a < p_2 < \cdots < p_{k-1} < b$:
\begin{equation}
p_i = a + \frac{i-1}{k-1} (b-a), 
\qquad
i = 1, \ldots, k.
\label{quants}
\end{equation}
This is a distribution-free method.

  \item {\em Uniformly-spaced quantiles\/} 
$F_*^{-1}(a) < F_*^{-1}(p_2) < \cdots < F_*^{-1}(p_{k-1}) < F_*^{-1}(b)$:
\begin{equation}
p_i = F_* \left( F_*^{-1}(a) + \frac{i-1}{k-1} 
\Big[ F_*^{-1}(b) - F_*^{-1}(a) \Big] \right), 
\qquad
i = 1, \ldots, k.
\label{uniQ}
\end{equation}
This is a distribution-dependent method.

  \item {\em Optimal} $a < p_2 < \cdots < p_{k-1} < b$. 
Optimal $p_i$'s are selected by numerically maximizing \eqref{AREg}:
\begin{equation}
(p_2^*, \ldots, p_{k-1}^*) ~=~
\operatorname*{arg\,max}_{a \, < p_2 < \cdots \, < p_{k-1} < \, b}
\mbox{ARE}_k \, \big( \mbox{log-gQLS}, \, \mbox{MLE} \big) ,
\qquad
k \geq 3.
\label{optP}
\end{equation}
This is a distribution-dependent method.
\end{enumerate}
The three strategies are compared in Table 2.4 for 
$(a, b) = (0.05, 0.95)$ and various $k$. From the table we first 
notice that the uniform spacing of $p_i$'s is optimal for log-Cauchy.
For the other distributions, it seems like \eqref{uniQ} selects
$p_i$'s somewhat closer to the optimal ones. However, choosing 
$k \geq 15$, makes $\mbox{ARE}_k$'s nearly identical across all 
strategies, and for each distribution considered here. Hence, 
based on this analysis and some additional explorations, we conclude 
that uniform selection of $p_1 < \cdots < p_k$ over $[a, b]$, 
with $0 < a \leq 0.05$, $0.95 \leq b < 1$, and $k \geq 15$, is 
a simple, distribution-free design which often yields very high 
AREs, and for log-Cauchy it is even the optimal design. (For 
the simulation studies, real data examples, and illustrations 
of this paper, we will use $k=25$.) 

\begin{center}
{\sc Table 2.4.} $p_i$ levels and $\mbox{ARE}_k$'s of 
log-gQLS of selected log-location-scale distributions
\\[-1ex]
under the strategies \eqref{quants}--\eqref{optP} with 
$(p_1 = a, p_k = b) = (0.05, 0.95)$ and $k = 7, 15, 25, 35, 45$.

\medskip

\begin{tabular}{|c|c|c|c|c|}
\hline
 & Log-Cauchy & Log-Laplace & Log-Logistic & Lognormal \\[-0.5ex]
 & \eqref{quants} \eqref{uniQ} \eqref{optP} &
   \eqref{quants} \eqref{uniQ} \eqref{optP} &
   \eqref{quants} \eqref{uniQ} \eqref{optP} &
   \eqref{quants} \eqref{uniQ} \eqref{optP} \\
\hline
\hline
$p_2$ & 0.200 ~0.074 ~0.200 & 0.200 ~0.108 ~0.110 & 
        0.200 ~0.123 ~0.142 & 0.200 ~0.136 ~0.127 \\[-1ex]
$p_3$ & 0.350 ~0.141 ~0.350 & 0.350 ~0.232 ~0.275 &
        0.350 ~0.273 ~0.290 & 0.350 ~0.292 ~0.268 \\[-1ex]
$p_4$ & 0.500 ~0.500 ~0.500 & 0.500 ~0.500 ~0.500 & 
        0.500 ~0.500 ~0.500 & 0.500 ~0.500 ~0.050 \\[-1ex]
$p_5$ & 0.650 ~0.859 ~0.650 & 0.650 ~0.768 ~0.725 &
        0.650 ~0.727 ~0.710 & 0.650 ~0.708 ~0.732 \\[-1ex]
$p_6$ & 0.800 ~0.926 ~0.800 & 0.800 ~0.892 ~0.890 &
        0.800 ~0.877 ~0.858 & 0.800 ~0.864 ~0.873 \\[-0.5ex]
$\mbox{ARE}_7$ & 0.934 ~0.728 ~0.934 & 0.919 ~0.926 ~0.927 & 
                 0.916 ~0.925 ~0.926 & 0.876 ~0.889 ~0.890 \\
\hline
$p_2$ & 0.114 ~0.058 ~0.114 & 0.114 ~0.069 ~0.094 & 
        0.114 ~0.074 ~0.083 & 0.114 ~0.079 ~0.077 \\[-1ex]
$p_3$ & 0.179 ~0.069 ~0.179 & 0.179 ~0.097 ~0.152 & 
        0.179 ~0.109 ~0.125 & 0.179 ~0.120 ~0.113 \\[-1ex]
$p_4$ & 0.243 ~0.086 ~0.243 & 0.243 ~0.134 ~0.214 & 
        0.243 ~0.157 ~0.178 & 0.243 ~0.174 ~0.161 \\[-1ex]
$p_5$ & 0.307 ~0.113 ~0.307 & 0.307 ~0.186 ~0.275 & 
        0.307 ~0.221 ~0.242 & 0.307 ~0.240 ~0.221 \\[-1ex]
$p_6$ & 0.371 ~0.161 ~0.371 & 0.371 ~0.259 ~0.334 & 
        0.371 ~0.301 ~0.318 & 0.371 ~0.319 ~0.298 \\[-1ex]
$p_7$ & 0.436 ~0.266 ~0.436 & 0.436 ~0.360 ~0.390 &
        0.436 ~0.396 ~0.405 & 0.436 ~0.407 ~0.392 \\[-1ex]
$p_8$ & 0.500 ~0.500 ~0.500 & 0.500 ~0.500 ~0.445 & 
        0.500 ~0.500 ~0.500 & 0.500 ~0.500 ~0.500 \\[-1ex]
$p_9$ & 0.564 ~0.734 ~0.564 & 0.564 ~0.640 ~0.500 & 
        0.564 ~0.604 ~0.595 & 0.564 ~0.593 ~0.608 \\[-1ex]
$p_{10}$ & 0.629 ~0.839 ~0.629 & 0.629 ~0.741 ~0.588 & 
           0.629 ~0.699 ~0.682 & 0.629 ~0.681 ~0.702 \\[-1ex]
$p_{11}$ & 0.693 ~0.887 ~0.693 & 0.693 ~0.814 ~0.673 & 
           0.693 ~0.779 ~0.758 & 0.693 ~0.760 ~0.779 \\[-1ex]
$p_{12}$ & 0.757 ~0.914 ~0.757 & 0.757 ~0.866 ~0.755 &
           0.757 ~0.843 ~0.822 & 0.757 ~0.826 ~0.839 \\[-1ex]
$p_{13}$ & 0.821 ~0.931 ~0.821 & 0.821 ~0.903 ~0.833 &
           0.821 ~0.891 ~0.875 & 0.821 ~0.880 ~0.887 \\[-1ex]
$p_{14}$ & 0.886 ~0.942 ~0.886 & 0.886 ~0.931 ~0.901 & 
           0.886 ~0.926 ~0.917 & 0.886 ~0.921 ~0.923 \\[-0.5ex]
$\mbox{ARE}_{15}$ & 0.987 ~0.913 ~0.987 & 0.943 ~0.944 ~0.944 & 
                    0.949 ~0.951 ~0.952 & 0.906 ~0.909 ~0.909 \\
\hline
$\mbox{ARE}_{25}$ & 0.995 ~0.966 ~0.995 & 0.947 ~0.947 ~0.947 &
                    0.955 ~0.956 ~0.956 & 0.911 ~0.912 ~0.912 \\
$\mbox{ARE}_{35}$ & 0.997 ~0.982 ~0.997 & 0.948 ~0.948 ~0.948 &
                    0.956 ~0.957 ~0.957 & 0.912 ~0.913 ~0.913 \\
$\mbox{ARE}_{45}$ & 0.998 ~0.989 ~0.998 & 0.948 ~0.948 ~0.948 &
                    0.957 ~0.957 ~0.957 & 0.913 ~0.913 ~0.913 \\
\hline
\end{tabular}
\end{center}

\bigskip

Finally, note that AREs of all log-gQLS estimators in this study 
are very high ($> 0.90$ when $k \geq 15$), which is remarkable. 
To put this in perspective, think  about the situation where 
$n=1000$ data points are observed and a log-location-scale loss 
model has to be fitted to the data at hand. While MLE would require 
all 1000 points to be explicitly included in its formulas, the 
log-gQLS estimator would rely on only 25 (or fewer) strategically 
selected points, yet achieve almost identical standard errors and 
offer performance protection against data contamination and model 
misspecification.

\subsection{Goodness of Fit}

For location-scale families, \citet[][]{ab25} constructed two 
goodness-of-fit tests using the 
$\widehat{\mbox{\boldmath $\beta$}}_{\mbox{\tiny gQLS}}$ 
estimator, which is more efficient than
$\widehat{\mbox{\boldmath $\beta$}}_{\mbox{\tiny oQLS}}$.
The first test was based on a quadratic form in model residuals
(which are defined as sample quantiles, i.e., those that were 
used in estimation of $\mbox{\boldmath $\beta$}$, minus 
model-estimated quantiles) and was called ``in-sample validation''. 
The main advantage of this test is that its distributional 
properties are analytically tractable and hence $p$-values 
can be easily computed. On the other hand, when $a$ and $1-b$ 
are increased, this test shrinks the range of quantiles which
are used to validate the model. Thus, subsequent comparisons of 
two model fits with different choices of $a$ and $b$ are not 
fair. Indeed, one can achieve a nearly perfect fit by choosing 
both $a$ and $b$ ($a < b$) around 0.50 and thus always win the 
``goodness-of-fit competition'' against the less extreme choices. 
To alleviate this problem, the second test proposed by the same 
authors was based on a combination of the model residuals and 
additional sample quantiles (the ones that had not been used 
for parameter estimation). It was called ``out-of-sample 
validation''. In this paper, we will modify the second test 
to make it workable for log-location-scale families.

To test the hypotheses 
\[
\left\{ 
\begin{array}{cl}
H_0: & X_1, \ldots, X_n 
\mbox{~ were generated by a log-location-scale family } F \\
H_A: & X_1, \ldots, X_n 
\mbox{~ were {\em not\/} generated by } F, \\
\end{array}
\right.
\]
the following procedure can be used.
\begin{enumerate}
  \item Select a model-validation set of sample log-quantiles:
$\mathbf{Y}_{\mbox{\tiny out}} = 
\left( \log \widehat{F}^{-1}(p_1^{\mbox{\tiny out}}), \ldots, 
\log \widehat{F}^{-1}(p_r^{\mbox{\tiny out}}) \right)'$, 
where 
$p_1^{\mbox{\tiny out}}, \ldots, p_r^{\mbox{\tiny out}}$ can be 
all different from or partially overlapping with $p_1, \ldots, p_k$ 
(recall that $p_1, \ldots, p_k$ were used for parameter estimation). 

  \item Using the linear regression model \eqref{l-reg}, predict 
the values of $\left( \log F^{-1}(p_1^{\mbox{\tiny out}}), 
\ldots, \log F^{-1}(p_r^{\mbox{\tiny out}}) \right)'$
with
$\mathbf{X_{\mbox{\tiny out}}} 
\widehat{\mbox{\boldmath $\beta$}}_{\mbox{\tiny log-gQLS}}$,
where
\[
\mathbf{X}_{\mbox{\tiny out}} ~=~ 
\begin{bmatrix}
1 & \cdot & \cdot & \cdot & 1 \\[1ex]
F_*^{-1}(p_1^{\mbox{\tiny out}}) & \cdot & \cdot & \cdot & 
F_*^{-1}(p_r^{\mbox{\tiny out}}) \\
\end{bmatrix}'
\]
and $\widehat{\mbox{\boldmath $\beta$}}_{\mbox{\tiny log-gQLS}} =
(\widehat{\mu}_{\mbox{\tiny log-gQLS}},\widehat{\sigma}_{\mbox{\tiny log-gQLS}})'$
is based on $\mathbf{Y} = 
\left( \log \widehat{F}^{-1}(p_1), \ldots, \log \widehat{F}^{-1}(p_k) \right)'$.

  \item Construct the following test statistic out of studentized residuals:
\begin{equation}
W_{\mbox{\tiny out}} ~=~ \frac{n}{\widehat{\sigma}^2_{\mbox{\tiny log-gQLS}}} 
\left( \mathbf{Y}_{\mbox{\tiny out}} - \mathbf{X}_{\mbox{\tiny out}} 
\widehat{\mbox{\boldmath $\beta$}}_{\mbox{\tiny log-gQLS}} \right)'
\mbox{\boldmath $\Sigma_{\mbox{\tiny out}}^{-1}$}
\left( \mathbf{Y}_{\mbox{\tiny out}} - \mathbf{X}_{\mbox{\tiny out}} 
\widehat{\mbox{\boldmath $\beta$}}_{\mbox{\tiny log-gQLS}} \right),
\label{gof2}
\end{equation}
where the elements of {\boldmath $\Sigma_{\mbox{\tiny out}}$} 
are given by $\sigma_{ij}^{\mbox{\tiny out}} = 
\frac{\min \{ p_i^{\mbox{\tiny out}} , \, p_j^{\mbox{\tiny out}} \} 
\, - \, p_i^{\mbox{\tiny out}} p_j^{\mbox{\tiny out}} }
{f_*(F_*^{-1}(p_i^{\mbox{\tiny out}})) 
f_*(F_*^{-1}(p_j^{\mbox{\tiny out}}))}$ ~for $~i, j = 1, \ldots, r$.

  \item Estimate the $p$-value associated with this test statistic
using the specified below bootstrap procedure. (We note in passing that 
theoretical treatment of the distribution of $W_{\mbox{\tiny out}}$ is 
an ongoing challenge and is beyond the scope of the current paper.)

\medskip

\begin{figure}[ht!]
\begin{minipage}{17cm}
\begin{enumerate}
  \item[] \underline{\hspace{\linewidth}}

\vspace{-1ex}

  \item[] {\bf ~ Bootstrap Procedure} (for finding 
the $p$-value of \eqref{gof2})

\vspace{-3ex}

  \item[] \underline{\hspace{\linewidth}}

\vspace{-1ex}

  \item[] {\bf Step 1.} Given the original sample, 
$X_1, \ldots, X_n$, the estimates of 
{\boldmath $\beta$} and $W_{\mbox{\tiny out}}$ 
are obtained. Denote them 
$\widehat{\mbox{\boldmath $\beta$}}^o_{\mbox{\tiny log-gQLS}} =
(\widehat{\mu}^o_{\mbox{\tiny log-gQLS}},
\widehat{\sigma}^o_{\mbox{\tiny log-gQLS}})'$
and $\widehat{W}^o_{\mbox{\tiny out}}$.
Remember that 
$\widehat{\mbox{\boldmath $\beta$}}^o_{\mbox{\tiny log-gQLS}}$
is computed using the quantile levels $p_1, \ldots, p_k$, 
while $\widehat{W}^o_{\mbox{\tiny out}}$ is based on
$p_1^{\mbox{\tiny out}}, \ldots, p_r^{\mbox{\tiny out}}$ and
$\widehat{\mbox{\boldmath $\beta$}}^o_{\mbox{\tiny log-gQLS}}$.

  \item[] {\bf Step 2.} Generate an {\em i.i.d\/}. sample 
  $X_1^{(b)}, \ldots, X_n^{(b)}$
from $F$ (assumed under $H_0$) using the parameter values
$\widehat{\mbox{\boldmath $\beta$}}^o_{\mbox{\tiny gQLS}} =
(\widehat{\mu}^o_{\mbox{\tiny gQLS}},
\widehat{\sigma}^o_{\mbox{\tiny gQLS}})'$.
Based on this sample, compute
$\widehat{\mbox{\boldmath $\beta$}}^{(b)}_{\mbox{\tiny log-gQLS}}$
(using $p_1, \ldots, p_k$) and
$\widehat{W}^{(b)}_{\mbox{\tiny out}}$ (using 
$p_1^{\mbox{\tiny out}}, \ldots, p_r^{\mbox{\tiny out}}$ and
$\widehat{\mbox{\boldmath $\beta$}}^{(b)}_{\mbox{\tiny log-gQLS}}$).

  \item[] {\bf Step 3.} Repeat {\sc Step 2} a $B$ number of times 
(e.g., $B=1000$) and save 
$\widehat{W}^{(1)}_{\mbox{\tiny out}}, \ldots, 
\widehat{W}^{(B)}_{\mbox{\tiny out}}$.

  \item[] {\bf Step 4.} Estimate the $p$-value of \eqref{gof2} by
\[
\widehat{p}_{\mbox{\tiny val}} ~=~ 
\frac{1}{B} \sum_{b=1}^B {\mbox{\large\bf 1}} 
\left\{ \widehat{W}^{(b)}_{\mbox{\tiny out}} > 
\widehat{W}^o_{\mbox{\tiny out}} \right\}
\]
and reject $H_0$ when 
$\widehat{p}_{\mbox{\tiny val}} \leq \alpha$
(e.g., $\alpha = 0.05$).

\vspace{-3ex}

  \item[] \underline{\hspace{\linewidth}}
\end{enumerate}
\end{minipage}
\end{figure}
\end{enumerate}

\subsection{Pareto and Weibull}

Pareto (type $I$) and Weibull are some of the most popular 
distributions for modeling claim severity. According to 
the inventory of continuous loss distributions 
\citep[][Appendix A]{kpw12}, their pdf, cdf and qf are 
parameterized as follows. For Pareto $I$,
\begin{eqnarray}
f_{\mbox{\tiny Pa}}(x) & = & 
\dfrac{\alpha \theta^{\alpha}}{x^{\alpha+1}}, \qquad x > \theta,
\nonumber
\\[1ex]
F_{\mbox{\tiny Pa}}(x) & = & 
1 - (\theta/x)^{\alpha}, \qquad x \geq \theta,
\nonumber
\\[1ex]
F^{-1}_{\mbox{\tiny Pa}}(u) & = & \theta (1-u)^{-1/\alpha}, 
\qquad 0 < u < 1.
\label{PaI}
\end{eqnarray}
And for Weibull,
\begin{eqnarray}
f_{\mbox{\tiny We}}(x) & = & 
\dfrac{\tau (x/\theta)^{\tau} e^{-(x/\theta)^{\tau}}}{x}, 
\qquad x > 0,
\nonumber
\\[1ex]
F_{\mbox{\tiny We}}(x) & = & 
1 - e^{-(x/\theta)^{\tau}}, \qquad x \geq 0,
\nonumber
\\[1ex]
F^{-1}_{\mbox{\tiny We}}(u) & = & 
\theta \big( - \log (1-u) \big)^{1/\tau}, \qquad 0 < u < 1.
\label{We}
\end{eqnarray}
It is clear from \eqref{PaI} and \eqref{We} that the logarithmic 
transformation of $F^{-1}$ converts Pareto $I$ and Weibull to 
{\em two-parameter\/} exponential and Gumbel ({\em minimum\/}), 
respectively, with: 
$\mu = \log \theta$, $\sigma = 1/\alpha$ (Pareto $I$) 
and 
$\mu = \log \theta$, $\sigma = 1/\tau$ (Weibull).
However, two-parameter exponential and Gumbel are skewed location-scale 
distributions, thus making Pareto $I$ and Weibull log-skewed location-scale 
families. Therefore, statistical inference for $\theta$, $\alpha$, and 
$\tau$ requires a bit more work than what was presented in Section 2.1.3. 
That is, after finding the log-oQLS and log-gQLS estimators of $\mu$ and 
$\sigma$, parameters $\theta$, $\alpha$, and $\tau$ are then estimated 
from the expressions $\theta = e^{\mu}$, $\alpha = 1/\sigma$, and 
$\tau = 1/\sigma$, respectively. And, to specify the ${\cal AN}$ 
distributions of such estimators, an additional application of 
the delta method is necessary. The following examples supply 
the missing details of this discussion.

\medskip

\noindent
{\bf Example 2.5} ~ Let $X_1, \ldots, X_n$ be i.i.d. random variables from 
$\mbox{Pareto} \, I \, (\theta, \alpha)$ with pdf and qf parameterized as 
in \eqref{PaI}. Suppose log-oQLS and log-gQLS of $\mu$ and $\sigma$ have 
been computed according to \eqref{oQLS} and \eqref{gQLS}, respectively. 
Then, their ${\cal AN}$ distributions follow \eqref{l-an-o} and 
\eqref{l-an-g}, respectively. Since $\theta = e^{\mu}$ and 
$\alpha = 1/\sigma$, the Jacobian of such transformations is given by
\[
\mathbf{D} ~=~ \begin{bmatrix}
\frac{\partial \theta}{\partial \mu} & \frac{\partial \theta}{\partial \sigma} 
\\[1ex]
\frac{\partial \alpha}{\partial \mu} & \frac{\partial \alpha}{\partial \sigma} 
\end{bmatrix} 
~=~
\begin{bmatrix}
e^{\mu} & 0 
\\[1ex]
0 & -\frac{1}{\sigma^2}
\end{bmatrix}
~=~
\begin{bmatrix}
\theta & 0 
\\[1ex]
0 & -\alpha^2
\end{bmatrix} .
\]
Therefore, it follows from the delta method that
$\left( 
\widehat{\theta}_{\mbox{\tiny log-oQLS}}, \,
\widehat{\alpha}_{\mbox{\tiny log-oQLS}}
\right)'
~=~
\left( 
e^{\widehat{\mu}_{\mbox{\tiny log-oQLS}}}, \,
\frac{1}{\widehat{\sigma}_{\mbox{\tiny log-oQLS}}}
\right)'$ is ${\cal AN}$ with the mean vector
$(\theta, \alpha)$ and $2 \times 2$ covariance-variance matrix
\[
\frac{\sigma^2}{n} \, \mathbf{D} 
\Big[ 
(\mathbf{X' X})^{-1} \mathbf{X'} 
\mbox{\boldmath $\Sigma_*$}
\mathbf{X} (\mathbf{X' X})^{-1}
\Big]
\mathbf{D'} 
~=~
\frac{1}{n} \,  \begin{bmatrix}
\theta/\alpha & 0 
\\[1ex]
0 & -\alpha
\end{bmatrix}
\Big[ 
(\mathbf{X' X})^{-1} \mathbf{X'} 
\mbox{\boldmath $\Sigma_*$}
\mathbf{X} (\mathbf{X' X})^{-1}
\Big]
\begin{bmatrix}
\theta/\alpha & 0 
\\[1ex]
0 & -\alpha
\end{bmatrix} .
\]

Further, following the same steps as for log-oQLS, 
$\left( 
\widehat{\theta}_{\mbox{\tiny log-gQLS}}, \,
\widehat{\alpha}_{\mbox{\tiny log-gQLS}}
\right)'
~=~
\left( 
e^{\widehat{\mu}_{\mbox{\tiny log-gQLS}}}, \,
\frac{1}{\widehat{\sigma}_{\mbox{\tiny log-gQLS}}}
\right)'$ is ${\cal AN}$ with the mean vector
$(\theta, \alpha)$ and $2 \times 2$ covariance-variance matrix
\[
\frac{\sigma^2}{n} \, \mathbf{D} 
\Big[ 
(\mathbf{X'} \mbox{\boldmath $\Sigma_*^{-1}$} \mathbf{X})^{-1}
\Big]
\mathbf{D'} 
~=~
\frac{1}{n} \,  \begin{bmatrix}
\theta/\alpha & 0 
\\[1ex]
0 & -\alpha
\end{bmatrix}
\Big[ 
(\mathbf{X'} \mbox{\boldmath $\Sigma_*^{-1}$} \mathbf{X})^{-1}
\Big]
\begin{bmatrix}
\theta/\alpha & 0 
\\[1ex]
0 & -\alpha
\end{bmatrix} .
\]

To enable the relative efficiency comparisons, the ${\cal AN}$ 
distribution of MLE has to be specified. Since parameter $\theta$ 
is part of the domain of the Pareto $I$ pdf, this distribution does 
not satisfy the well-known regularity conditions 
\citep[see][Section 4.2.2]{s02a}. Therefore, only the univariate 
case, when $\theta$ is assumed known, can be explored. 
But this is a straightforward exercise, as MLE-related results 
are readily available \citep[see, e.g.,][Chapter 5]{a14}:
$\widehat{\alpha}_{\mbox{\tiny MLE}} ~=~ 
n \big/ \sum_{i=1}^n \log (X_i/\theta)$, 
the Fisher information matrix (scalar) is $\mathbf{I} = 
\alpha^{-2} \mathbf{I_*} = \alpha^{-2}$, and 
$\widehat{\alpha}_{\mbox{\tiny MLE}}$ is ${\cal AN}
\left( \alpha, \, \alpha^2/n \right)$.

Also, according to Note 2.2, when $\theta$ is assumed known, 
log-oQLS and log-gQLS formulas need a few modifications. 
In particular, the modified data vector is given by
\[
\mathbf{Y} ~=~ 
\left( \log \big( \widehat{F}^{-1}(p_1) \big) - \mu, \ldots, \, 
\log \big( \widehat{F}^{-1}(p_k) \big) - \mu \right)' ~=~
\left( \log \big( \widehat{F}^{-1}(p_1)/\theta \big), \ldots, \, 
\log \big( \widehat{F}^{-1}(p_k)/\theta \big) \right)',
\]
and the design matrix (vector) becomes
\[
\mathbf{X} ~=~ \left[ F_*^{-1}(p_1), \ldots, \, F_*^{-1}(p_k) \right]'
~=~ \left[ -\log (1-p_1), \ldots, \, -\log (1-p_k) \right]'.
\]
Therefore,
\[
\widehat{\alpha}_{\mbox{\tiny log-oQLS}} ~=~ \frac{1}
{ ( \mathbf{X'} \mathbf{X} )^{-1} \mathbf{X'} \mathbf{Y} }
\quad \mbox{is} \quad
{\cal AN} \left( \alpha, \, \frac{\alpha^2}{n} \, 
(\mathbf{X' X})^{-1} \mathbf{X'} \mbox{\boldmath $\Sigma_*$} 
\mathbf{X} (\mathbf{X' X})^{-1}
\right)
\]
and
\[
\widehat{\alpha}_{\mbox{\tiny log-gQLS}} ~=~ \frac{1}
{ ( \mathbf{X'} \mbox{\boldmath $\Sigma_*^{-1}$} \mathbf{X} )^{-1} 
\mathbf{X'} \mbox{\boldmath $\Sigma_*^{-1}$} \mathbf{Y} }
\quad \mbox{is} \quad
{\cal AN} \left( \alpha, \, \frac{\alpha^2}{n} \, 
(\mathbf{X'} \mbox{\boldmath $\Sigma_*^{-1}$} \mathbf{X})^{-1} 
\right).
\]

\medskip

Finally, the ARE formulas are given by
\[
\mbox{ARE} \, \big( \mbox{log-oQLS}, \, \mbox{MLE} \big) ~=~
\left( \frac{\mbox{det} \left[ \dfrac{\alpha^2}{n} \right]}
{\mbox{det} \left[ \dfrac{\alpha^2}{n} \, \Big[ 
(\mathbf{X' X})^{-1} \mathbf{X'} \mbox{\boldmath $\Sigma_*$}
\mathbf{X} (\mathbf{X' X})^{-1} \Big] \right]} 
\right)^{1/1}
=~
\frac{1}{(\mathbf{X' X})^{-1} \mathbf{X'} 
\mbox{\boldmath $\Sigma_*$} \mathbf{X} (\mathbf{X' X})^{-1}}
\]
and
\[
\mbox{ARE} \, \big( \mbox{log-gQLS}, \, \mbox{MLE} \big) ~=~
\left( \frac{\mbox{det} \left[ \dfrac{\alpha^2}{n} \right]}
{\mbox{det} \left[ \dfrac{\alpha^2}{n} \, \Big[ 
(\mathbf{X'} \mbox{\boldmath $\Sigma_*^{-1}$} \mathbf{X})^{-1} 
\Big] \right]} 
\right)^{1/1} 
=~
\frac{1}{(\mathbf{X'} \mbox{\boldmath $\Sigma_*^{-1}$} \mathbf{X})^{-1}}
\]
For numerical illustrations, see Table 2.5.
\hfill $\Box$

\bigskip

\noindent
{\bf Example 2.6} ~ Let $X_1, \ldots, X_n$ be i.i.d. random variables 
from $\mbox{Weibull} \, (\theta, \tau)$ with pdf and qf parameterized 
as in \eqref{We}. Suppose log-oQLS and log-gQLS of $\mu$ and $\sigma$ 
have been computed according to \eqref{oQLS} and \eqref{gQLS}, 
respectively. Then, their ${\cal AN}$ distributions follow \eqref{l-an-o} 
and \eqref{l-an-g}, respectively. Since the transformations 
$\theta = e^{\mu}$ and $\tau = 1/\sigma$ are the same as in Example 2.5, 
the ${\cal AN}$ distributions of log-oQLS and log-gQLS follow immediately:
$\left( 
\widehat{\theta}_{\mbox{\tiny log-oQLS}}, \,
\widehat{\tau}_{\mbox{\tiny log-oQLS}}
\right)'
~=~
\left( 
e^{\widehat{\mu}_{\mbox{\tiny log-oQLS}}}, \,
\frac{1}{\widehat{\sigma}_{\mbox{\tiny log-oQLS}}}
\right)'$ is ${\cal AN}$ with the mean vector
$(\theta, \tau)$ and $2 \times 2$ covariance-variance matrix
\[
\frac{\sigma^2}{n} \, \mathbf{D} 
\Big[ 
(\mathbf{X' X})^{-1} \mathbf{X'} 
\mbox{\boldmath $\Sigma_*$}
\mathbf{X} (\mathbf{X' X})^{-1}
\Big]
\mathbf{D'} 
~=~
\frac{1}{n} \,  \begin{bmatrix}
\theta/\tau & 0 
\\[1ex]
0 & -\tau
\end{bmatrix}
\Big[ 
(\mathbf{X' X})^{-1} \mathbf{X'} 
\mbox{\boldmath $\Sigma_*$}
\mathbf{X} (\mathbf{X' X})^{-1}
\Big]
\begin{bmatrix}
\theta/\tau & 0 
\\[1ex]
0 & -\tau
\end{bmatrix} 
\]
and
$\left( 
\widehat{\theta}_{\mbox{\tiny log-gQLS}}, \,
\widehat{\tau}_{\mbox{\tiny log-gQLS}}
\right)'
~=~
\left( 
e^{\widehat{\mu}_{\mbox{\tiny log-gQLS}}}, \,
\frac{1}{\widehat{\sigma}_{\mbox{\tiny log-gQLS}}}
\right)'$ is ${\cal AN}$ with mean $(\theta, \tau)$ and 
$2 \times 2$ covariance-variance matrix
\[
\frac{\sigma^2}{n} \, \mathbf{D} 
\Big[ 
(\mathbf{X'} \mbox{\boldmath $\Sigma_*^{-1}$} \mathbf{X})^{-1}
\Big]
\mathbf{D'} 
~=~
\frac{1}{n} \,  \begin{bmatrix}
\theta/\tau & 0 
\\[1ex]
0 & -\tau
\end{bmatrix}
\Big[ 
(\mathbf{X'} \mbox{\boldmath $\Sigma_*^{-1}$} \mathbf{X})^{-1}
\Big]
\begin{bmatrix}
\theta/\tau & 0 
\\[1ex]
0 & -\tau
\end{bmatrix} .
\]

Further, the ${\cal AN}$ distribution of MLE can be derived by 
exploiting the relationship between the Weibull and Gumbel 
(minimum) distributions. For Gumbel (minimum), the Fisher 
information matrix is given by
\[
\mathbf{I} ~=~ \sigma^{-2} \mathbf{I_*} ~=~
\sigma^{-2}
\begin{bmatrix}
1 & 1 - \gamma \\
1 - \gamma & \frac{\pi^2}{6} + (1-\gamma)^2
\end{bmatrix} ,
\]
where $\gamma \approx 0.57721$ is the Euler-Mascheroni 
constant. Then, it follows from the delta method that 
$\left( 
\widehat{\theta}_{\mbox{\tiny MLE}}, \,
\widehat{\tau}_{\mbox{\tiny MLE}}
\right)'
~=~
\left( 
e^{\widehat{\mu}_{\mbox{\tiny MLE}}}, \,
\frac{1}{\widehat{\sigma}_{\mbox{\tiny MLE}}}
\right)'$ is ${\cal AN}$ with mean $(\theta, \tau)$ and 
$2 \times 2$ covariance-variance matrix
\[
\frac{\sigma^2}{n} \, \mathbf{D} 
\Big[ 
\mathbf{I_*^{-1}}
\Big]
\mathbf{D'} 
~=~
\frac{1}{n} \,  \begin{bmatrix}
\theta/\tau & 0 
\\[1ex]
0 & -\tau
\end{bmatrix}
\Big[ 
\mathbf{I_*^{-1}}
\Big]
\begin{bmatrix}
\theta/\tau & 0 
\\[1ex]
0 & -\tau
\end{bmatrix} .
\]

Finally, for computation of AREs, the formulas \eqref{AREo} and 
\eqref{AREg} still remain valid, which is due to the properties 
of the matrix determinant:
\begin{eqnarray*}
\mbox{ARE} \, \big( \mbox{log-oQLS}, \, \mbox{MLE} \big) & = &
\left( \frac{\mbox{det} \left[ 
\dfrac{\sigma^2}{n} \, \mathbf{D} \Big[ \mathbf I_*^{-1} \Big] \mathbf{D'} 
\right]}{\mbox{det} 
\left[ \dfrac{\sigma^2}{n} \, \mathbf{D} \Big[ 
(\mathbf{X' X})^{-1} \mathbf{X'} \mbox{\boldmath $\Sigma_*$}
\mathbf{X} (\mathbf{X' X})^{-1} \Big] 
\mathbf{D'} \right]} 
\right)^{1/2}
\\[1.5ex]
 & = &
\left( \frac{\mbox{det} \left[ \mathbf I_*^{-1} \right]}
{\mbox{det} 
\big[
(\mathbf{X' X})^{-1} \mathbf{X'} 
\mbox{\boldmath $\Sigma_*$}
\mathbf{X} (\mathbf{X' X})^{-1}
\big]} 
\right)^{1/2}
\end{eqnarray*}
and
\[
\mbox{ARE} \, \big( \mbox{log-gQLS}, \, \mbox{MLE} \big) ~=~
\left( \frac{\mbox{det} \left[ \dfrac{\sigma^2}{n} \,
\mathbf{D} \Big[ \mathbf I_*^{-1} \Big] \mathbf{D'} \right]}
{\mbox{det} 
\left[ \dfrac{\sigma^2}{n} \, \mathbf{D} \Big[
( \mathbf{X'} \mbox{\boldmath $\Sigma^{-1}_*$} \mathbf{X} )^{-1} 
\Big] \mathbf{D'}
\right]} 
\right)^{1/2} 
 =~
\left( \frac{\mbox{det} \left[ \mathbf I_*^{-1} \right]}
{\mbox{det} 
\big[
( \mathbf{X'} \mbox{\boldmath $\Sigma^{-1}_*$} \mathbf{X} )^{-1}
\big]} 
\right)^{1/2} .
\]
For numerical illustrations, see Table 2.5.
\hfill $\Box$

\medskip

Using the formulas of Examples 2.5--2.6, in Table 2.5 we provide ARE 
values of log-oQLS and log-gQLS for the parameters of Pareto $I$ and 
Weibull distributions.

\begin{center}
{\sc Table 2.5.} AREs of log-oQLS and log-gQLS for Pareto $I$ 
and Weibull distributions.
\\[-1ex]
The quantiles are selected according to \eqref{quants} with 
$(a, b) = (0.05, 0.95)$ and various $k$.

\medskip

\begin{tabular}{|c|c|ccc|ccc|ccc|}
\hline
Probability & Estimation & \multicolumn{9}{|c|}{$k$} \\[-0.5ex]
\cline{3-11}
Distribution & Method & 2 & 3 & 4 & 5 & 7 & 9 & 10 & 15 & 25 \\
\hline
\hline
Pareto $I$ & log-oQLS & 0.473 & 0.511 & 0.566 & 0.621 & 0.712 & 
0.774 & 0.797 & 0.858 & 0.891 \\
 & log-gQLS & 0.508 & 0.779 & 0.858 & 0.892 & 0.921 & 0.933 & 
0.936 & 0.944 & 0.948 \\
\hline
Weibull & log-oQLS & 0.488 & 0.572 & 0.611 & 0.637 & 0.677 & 
0.704 & 0.714 & 0.742 & 0.757 \\
 & log-gQLS & 0.488 & 0.714 & 0.791 & 0.828 & 0.861 & 0.874 & 
0.879 & 0.888 & 0.893 \\
\hline
\end{tabular}
\end{center}

\medskip

\section{Synthetic Data Illustrations}

In this section, we conduct a simulation study with the objective 
of verifying and augmenting the theoretical properties established in 
Section 2. We start by describing the study design in Section 3.1. 
Then we explore how the MLE, log-oQLS and log-gQLS estimators 
perform when data are ``clean'' and when they are contaminated 
by outliers (Section 3.2). To see how effective the robust 
estimators are in identifying outliers, we apply a few 
well-established outlier-labeling rules on a simulated data set 
(Section 3.3). We finish the study by investigating the power 
of the goodness-of-fit test, based on the statistic \eqref{gof2}, 
against several alternatives (Section 3.4).

\subsection{Study Design}

The study design is based on the following choices.
\begin{figure}[ht!]
\begin{minipage}{17cm}
\begin{enumerate}
  \item[] \underline{\hspace{\linewidth}}

\vspace{-1ex}

  \item[] {\bf ~ Simulation Design}

\vspace{-3ex}

  \item[] \underline{\hspace{\linewidth}}

\vspace{-1ex}

\begin{itemize}
  \item {\em Data-generating distributions\/} (i.e., the contaminated 
model $f_{\varepsilon} = (1-\varepsilon) \, f_0 + \varepsilon \, f_1$).
     \begin{itemize}
       \item[$*$] {\em Lognormal\/} (LN). ~$f_0$ is 
LN$\, (\mu_0 = 0, \sigma_0 = 1)$ and 
$f_1$ is LN$\, (\mu_1 = 2, \sigma_1 = 2)$.

       \item[$*$] {\em Weibull\/} (We). ~$f_0$ is 
We$\, (\theta_0 = 5, \tau_0 = 3/4)$ and 
$f_1$ is We$\, (\theta_1 = 5, \tau_1 = 1/4)$.

       \item[$*$] {\em Levels of contamination\/}.
~$\varepsilon = 0, \, 0.03, \, 0.08$.
     \end{itemize}

  \item {\em Estimators\/}. ~MLE, log-oQLS (abbreviated `o') and 
log-gQLS (abbreviated `g'). For `o' and `g' estimators, the 
quantiles are selected according to \eqref{quants} with
$(a_1, b_1) = (0.02, 0.98)$,
$(a_2, b_2) = (0.05, 0.95)$,
$(a_3, b_3) = (0.10, 0.90)$
and $k = 25$ (in all cases).
 
  \item {\em Goodness-of-fit test\/} (at $\alpha = 0.05$). 
~Based on $W_{\mbox{\tiny out}}$, given by \eqref{gof2}.

  \item {\em Quantile levels for model validation\/}. 
~$p_1^{\mbox{\tiny out}} = 0.01, \, p_2^{\mbox{\tiny out}} = 0.03, \ldots, 
\, p_{49}^{\mbox{\tiny out}} = 0.97, \, p_{50}^{\mbox{\tiny out}} = 0.99$.
  
  \item {\em Sample sizes\/}. ~$n = 10^2, \, 500, \, 10^3$.
  
  \item {\em Number of bootstrap samples\/}. ~$B = 10^3$.

  \item {\em Number of Monte Carlo runs\/}. ~$M = 10^4$.  
\end{itemize}

\vspace{-4ex}

  \item[] \underline{\hspace{\linewidth}}

\end{enumerate}
\end{minipage}
\end{figure}

In Section 3.2, for any given distribution, we generate $M = 10^4$ 
random samples of size $n = 500$. For each sample, we estimate 
$\mu$ and $\sigma$ (or $\theta$ and $\tau$) of the distribution 
using the MLE, log-oQLS, and log-gQLS estimators. The results 
are then presented using boxplots.
In Section 3.3, we generate a $8\%$ contaminated Weibull sample 
of size $n = 500$. After applying several outlier-labeling rules
we evaluate their success and failure rates. In Section 3.4, the 
data are generated similarly as in Section 3.2, but $n=100$ and 
$1000$. For each sample, we assume several plausible distributions 
and monitor the $H_0$ rejection rates of the goodness-of-fit test 
\eqref{gof2}. The results are summarized using tables.

\subsection{Data Contamination}

To start with, in Figure 3.1 we plot the clean and contaminated 
density functions of the lognormal and Weibull distributions. It is 
clear from the plots that for each model, $f_0$, $f_{0.03}$, and 
$f_{0.08}$ are practically indistinguishable, as all three curves 
overlap almost everywhere including the tails. This implies that 
standard diagnostic tools (e.g., histogram or quantile-quantile 
plot) would be of little help in identifying which of these models 
generated the data, hence, assuming $f_0$ is a reasonable way forward.
However, as our investigations demonstrate, the parameter estimates 
can move substantially away from their targets if data are generated 
by $f_{0.03}$ or $f_{0.08}$, instead of the assumed $f_0$.

\medskip

\begin{center}
\resizebox{170mm}{65mm}{\includegraphics{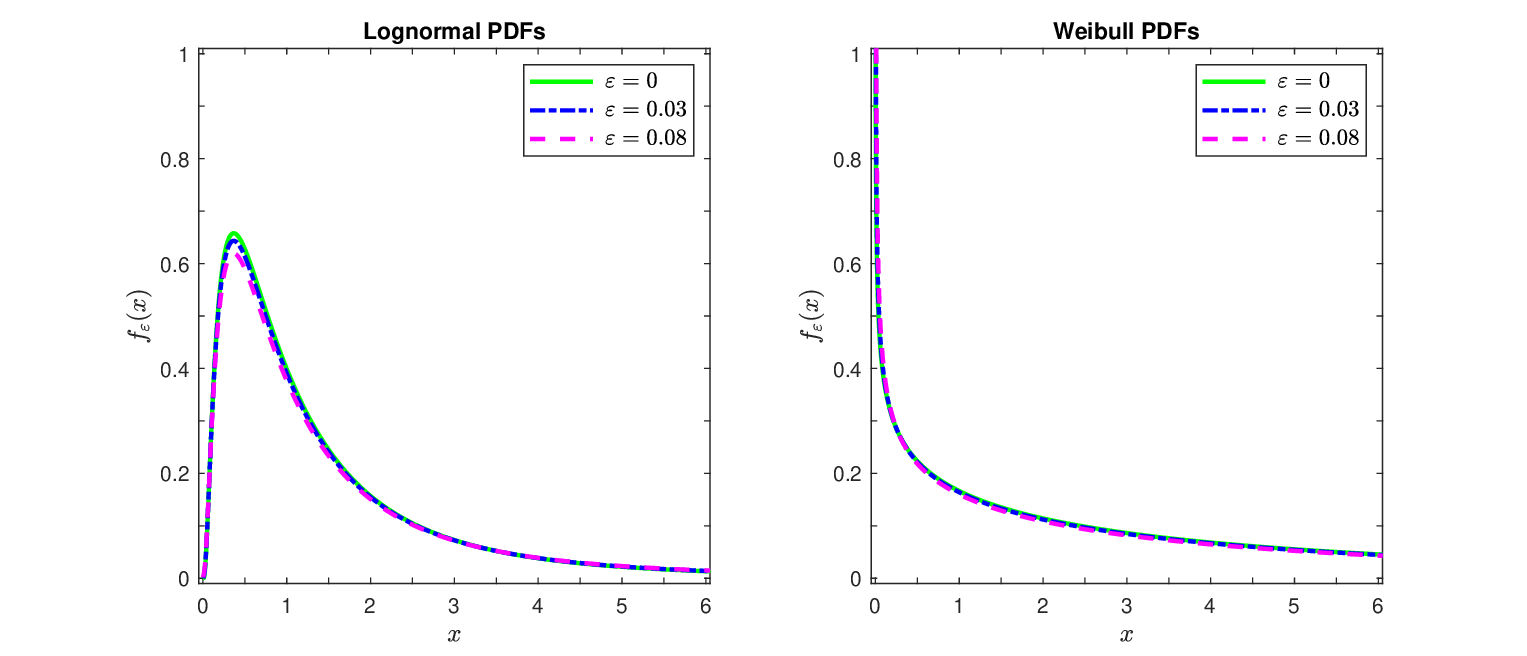}}
\\[-1ex]
{\sc Figure 3.1.} ~Lognormal and Weibull PDFs, $f_{\varepsilon} = 
(1-\varepsilon) f_0 + \varepsilon f_1$, for various choices of 
$\varepsilon$: 
\\[-1ex]
$\varepsilon = 0$ (no contamination),
$\varepsilon = 0.03$ (light contamination),
$\varepsilon = 0.08$ (moderate contamination).
\\[-1ex]
For the lognormal densities (LN): $f_0$ is 
LN$\, (\mu_0 = 0, \sigma_0 = 1)$ and 
$f_1$ is LN$\, (\mu_1 = 2, \sigma_1 = 2)$.
\\[-1ex]
For the Weibull densities (We): $f_0$ is 
We$\, (\theta_0 = 5, \tau_0 = 3/4)$ and 
$f_1$ is We$\, (\theta_1 = 5, \tau_1 = 1/4)$.
\end{center}

\medskip

\noindent
Next, we present boxplots of 10,000 estimates of $\mu$ and $\sigma$ 
(for lognormal in Figure 3.2) and $\theta$ and $\tau$ (for Weibull in
Figure 3.3) estimators. When data are clean, we see that all estimators 
cluster around their respective targets, with MLE performing slightly 
better than log-oQLS and log-gQLS in estimating $\sigma$ and $\tau$. 
However, when data are contaminated ($\varepsilon > 0$), all estimators 
get pulled away from their targets toward the parameter values of the 
contaminating model $f_1$. As expected for such scenarios, robust 
estimators are less affected by contamination while non-robust MLE
exhibits large bias and inflated variability. Interestingly, for 
the Weibull distribution, only parameter $\tau$ was contaminated, 
but MLE of $\theta$ got distorted too, whereas robust log-oQLS 
and log-gQLS (for $\theta$) stayed spot on.

\medskip

\begin{center}
\resizebox{170mm}{90mm}{\includegraphics{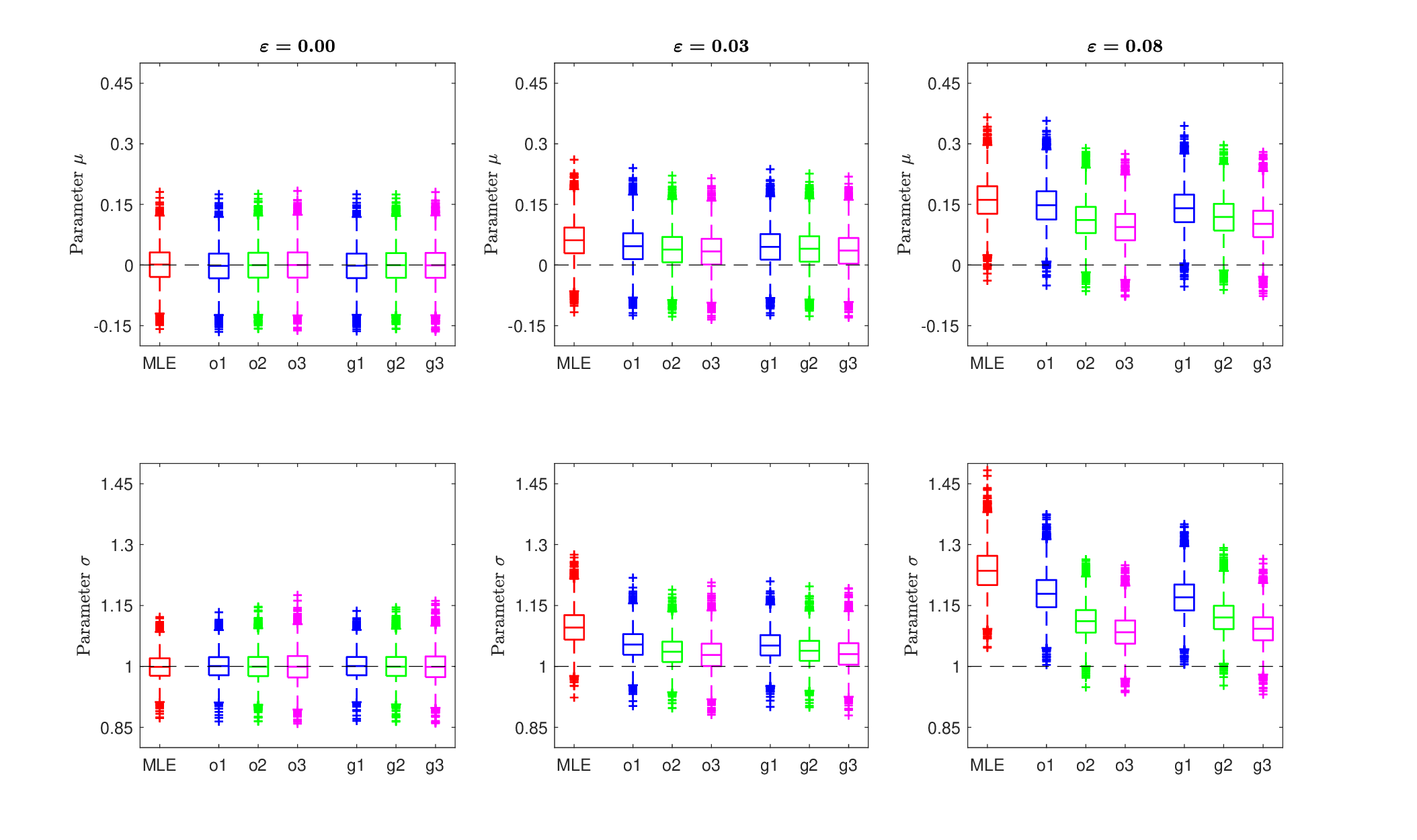}}
\\[-1ex]
{\sc Figure 3.2.} ~Lognormal$\, (\mu, \sigma)$ distributions. 
Boxplots of $\widehat{\mu}$ and $\widehat{\sigma}$ under clean 
($\varepsilon = 0.00$) and 
\\[-1ex]
contaminated ($\varepsilon = 0.03, \, 0.08$) data scenarios, 
using MLE, log-oQLS/log-gQLS (`o'/`g') estimators. 
\\[-1ex]
Here $(a, b)$ is equal to 
$(0.02, 0.98)$ for o1/g1, 
$(0.05, 0.95)$ for o2/g2, 
and $(0.10, 0.90)$ for o3/g3.
\end{center}

\medskip

\begin{center}
\resizebox{170mm}{90mm}{\includegraphics{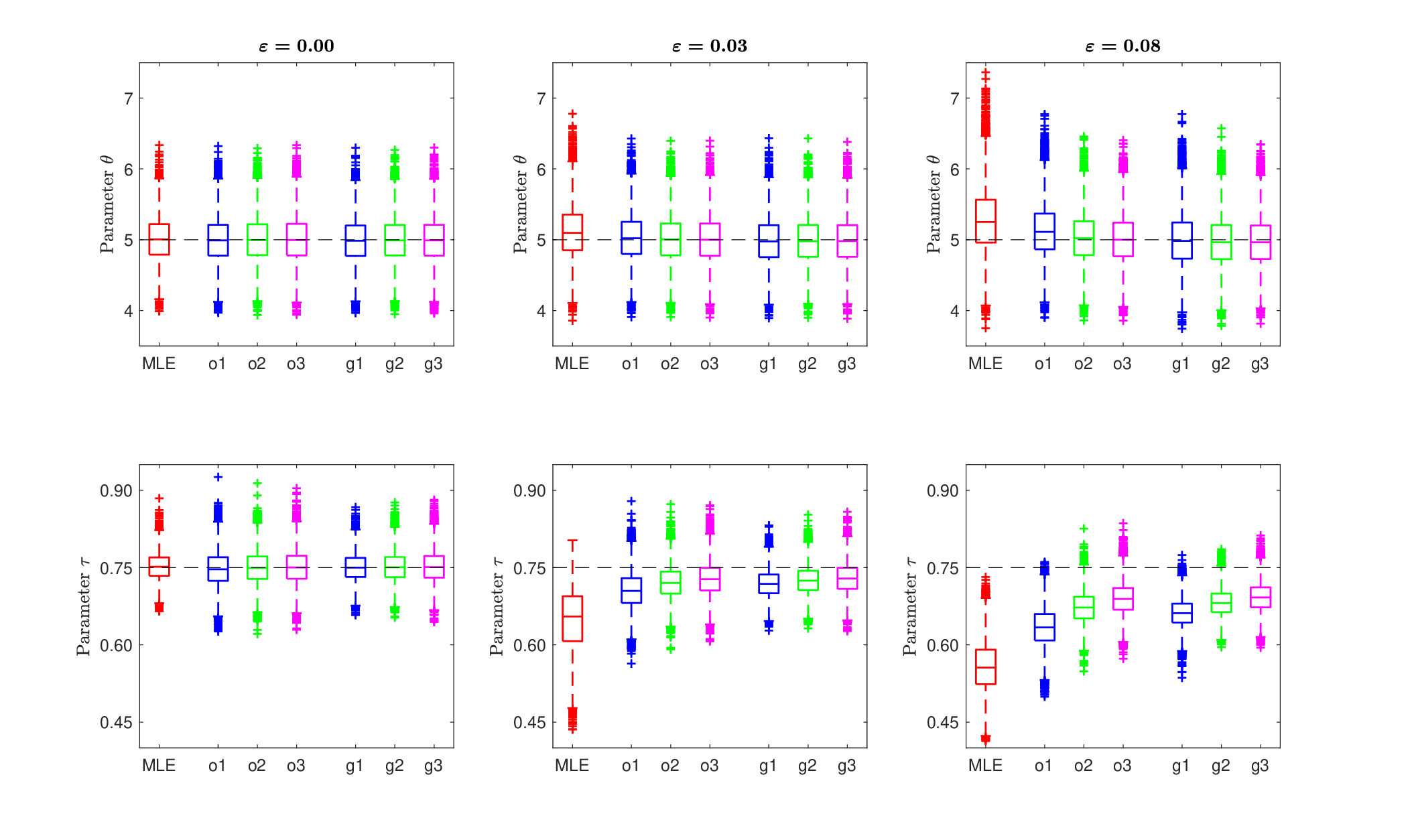}}
\\[-1ex]
{\sc Figure 3.3.} ~Weibull$\, (\theta, \tau)$ distributions. 
Boxplots of $\widehat{\theta}$ and $\widehat{\tau}$ under clean 
($\varepsilon = 0.00$) and 
\\[-1ex]
contaminated ($\varepsilon = 0.03, \, 0.08$) data scenarios, 
using MLE, log-oQLS/log-gQLS (`o'/`g') estimators. 
\\[-1ex]
Here $(a, b)$ is equal to 
$(0.02, 0.98)$ for o1/g1, 
$(0.05, 0.95)$ for o2/g2, 
and $(0.10, 0.90)$ for o3/g3.
\end{center}

\medskip

\subsection{Identification of Outliers}

To see how nonparametric and parametric outlier-labeling rules work, 
we generated 460 observations from a clean Weibull distribution ($f_0$) 
and 40 observations from a contaminating Weibull distribution ($f_1$). 
Then, the combined sample of size $n = 500$ was treated as a realization 
of an $8\%$ contaminated Weibull distribution ($f_{0.08}$). Summary 
statistics for each sample are provided in Table 3.1.

\begin{center}
{\sc Table 3.1.} Summary statistics of the clean ($f_0$), 
contaminating ($f_1$), and $8\%$ contaminated
\\[-1ex]
($f_{0.08}$) Weibull data, where 
$f_0$ is We$\, (\theta_0 = 5, \tau_0 = 3/4)$ and 
$f_1$ is We$\, (\theta_1 = 5, \tau_1 = 1/4)$.

\medskip

\begin{tabular}{|c|c|ccccc|cc|}
\hline
Data & Sample & 
\multicolumn{5}{|c|}{Five-Number Summary} &
\multicolumn{2}{|c|}{Moments} \\[-0.5ex]
Generated by & Size & 
$min$ & $q_1$ & $q_2$ & $q_3$ & $max$ & $mean$ & $std. \, dev.$ \\
\hline
\hline
$f_0$ & 460 & $4.34 \cdot 10^{-3}$ & 1.04 & 3.14 &  8.78 & $5.44 \cdot 10^{1}$ & 
6.21 & 8.11 \\ 
$f_1$ &  40 & 
$9.20 \cdot 10^{-9}$ & 0.01 & 0.19 & 11.67 & $5.34 \cdot 10^{3}$ & 232 & 880 \\
$f_{0.08}$ & 500 & 
$9.20 \cdot 10^{-9}$ & 0.77 & 3.05 & 8.78 & $5.34 \cdot 10^{3}$ & 24 & 254 \\
\hline
\end{tabular}
\end{center}

\medskip

Clearly, the contaminating sample is more left-skewed than the clean one 
because the minimum and the first two quartiles of $f_1$ are significantly 
smaller than those of $f_0$. On the other hand, $f_1$ also has {\em larger\/} 
data points in the right tail than $f_0$; this can be inferred from the values 
of $q_3$, $max$, $mean$, and $std. \, dev.$ Now, let us examine those extreme 
tails more carefully. In Table 3.2, ten most extreme observations in each tail 
are listed for all three samples.

\begin{center}
{\sc Table 3.2.} Sample extremes of the clean ($f_0$), 
contaminating ($f_1$), and $8\%$ contaminated
\\[-1ex]
($f_{0.08}$) Weibull data, where 
$f_0$ is We$\, (\theta_0 = 5, \tau_0 = 3/4)$ and 
$f_1$ is We$\, (\theta_1 = 5, \tau_1 = 1/4)$.

\medskip

\begin{tabular}{|c|cc|cc|cc|cc|cc|}
\hline
Data & \multicolumn{10}{|c|}{Most Extreme Data} \\[-0.5ex]
Generated by & \multicolumn{10}{|c|}{} \\ 
\hline
\multicolumn{11}{l}{{\em 10 Lowest Observations\/} {\footnotesize $(\times 10^{-3})$}} \\ 
\hline
$f_0$ & \fbox{4.34} & \fbox{5.06} & 5.52 & 6.14 & 7.62 & 10.06 & 
10.74 & 16.11 & 26.96 & 38.82 \\
$f_1$ & $9.20 \cdot 10^{-6}$ & 
0.01 & 0.11 & 0.52 & 1.22 & 2.80 & 3.39 & 4.67 & 5.59 & 7.71 \\
$f_{0.08}$ & $9.20 \cdot 10^{-6}$ & 
0.01 & 0.11 & 0.52 & 1.22 & 2.80 & 3.39 & \fbox{4.34} & 4.67 & \fbox{5.06} \\
\hline
\multicolumn{11}{l}{\em 10 Highest Observations} \\ 
\hline
$f_0$ & 34.45 & 36.37 & 36.96 & 37.16 & 38.84 & 40.94 & 48.05 & 
48.92 & 51.10 & \fbox{54.42} \\
$f_1$ & 17.5 & 73.4 & 137 & 180 & 237 & 239 & 694 & 734 & 1640 & 5340 \\ 
$f_{0.08}$ & \fbox{54.42} & 73.4 & 137 & 180 & 237 & 239 & 694 & 734 & 
1640 & 5340 \\
\hline
\end{tabular}
\end{center}

\medskip

Looking at these data it is natural to ask: What is an outlier? We will 
rely on the following interpretation: The data point that is smaller than 
the minimum or larger than the maximum observation from $f_0$ is an outlier. 
For example, as can be seen from Table 3.2 (lines $f_{0.08}$; the boxed 
entries highlight those $f_0$ extremes that appear as extremes within 
the $f_{0.08}$ sample), all observations below $4.34 \times 10^{-3}$ 
and above $54.4$ were generated by $f_1$ and are indeed outliers. In 
the generated sample of 500 observations, there are 7 {\em lower\/} and 
9 {\em upper\/} outliers. The remaining 24 observations from $f_1$ fall 
between the two extreme thresholds and assimilate with the data from 
$f_0$, making it impossible to distinguish them from genuine data. 
Indeed, after observing the sample from $f_{0.08}$, how can one tell 
that $4.67 \times 10^{-3}$, which falls between $4.34 \times 10^{-3}$ 
and $5.06 \times 10^{-3}$, is generated by $f_1$ and not $f_0$? 
Fortunately, the assimilated data are also harmless, i.e., their 
influence on the estimates of $f_0$ parameters is similar to that 
from the $f_0$ data. In view of this discussion, the quality of an 
outlier-labeling rule will be assessed by measuring its success rates 
in identifying the 7 lower and 9 upper outliers (high rates preferred), 
as well as the rate of outlier-declarations in the non-extreme region. 
Since the latter decisions are very likely to include `false positives', 
a low rate of declared outliers in this region is preferred.

There are many outlier-labeling rules available in the literature. 
For illustrative purposes, we will select two nonparametric and two 
parametric rules.
\begin{itemize}
  \item {\em Nonparametric rules\/}. The classic approach of 
\citet[][Section 2]{t77} is a very popular method, according to which 
outliers are those data points that fall outside the following interval:
\begin{equation}
\big[ q_1 - K \cdot (q_3-q_1); \, q_3 + K \cdot (q_3-q_1) \big],
\label{tukey}
\end{equation}
where $q_1$ and $q_3$ denote the first and third sample quartile, 
respectively, and $K=1.5$ (sometimes $K=3$ is used). The rule 
\eqref{tukey} works quite well for symmetric distributions. To make 
it better suited for asymmetrically-distributed data, \citet{k90} 
proposed the following modification:
\begin{equation}
\big[ q_1 - K \cdot (q_2-q_1); \, q_3 + K \cdot (q_3-q_2) \big],
\label{kimber}
\end{equation}
where $K=1.5$ and $q_2$ is the sample median.

  \item {\em Parametric rules\/}. The parametric approach of 
\citet{bi07} is designed for right-skewed distributions and focuses 
only on the upper outliers. It starts with the right end-point of 
\eqref{tukey} and replaces $q_1$ and $q_3$ with their respective 
parametric estimates, $\widehat{q}_1 = \widehat{F}^{-1}(1/4)$ and 
$\widehat{q}_3 = \widehat{F}^{-1}(3/4)$. 
It also uses the distribution of the largest order statistic, 
$X_{(n)}$, to estimate the number of the interquartile steps 
$\widehat{K} = 
\frac{ \widehat{F}^{-1}((1-\alpha)^{1/n}) - 
\widehat{q}_3}{\widehat{q}_3-\widehat{q}_1}$.
This, in turn, transforms \eqref{tukey} into:
\begin{equation}
\big( -\infty; \, \widehat{q}_3 + \widehat{K} \cdot
(\widehat{q}_3 - \widehat{q}_1) \big] ~=~ 
\big( -\infty; \, \widehat{F}^{-1} \big( (1-\alpha)^{1/n} \big) \big],
\label{iglewicz1}
\end{equation}
where, depending upon the context, $-\infty$ can be replaced by 0 
or some other finite (known) lower bound. The term 
$F^{-1} \big( (1-\alpha)^{1/n} \big)$ is derived by solving 
$\alpha = \mbox{\bf P} \big[ X_{(n)} > t \big] = 1 - [ F(t) ]^n$
for $t$. Note that this outlier-screening rule is dependent on 
the sample size $n$, making it more flexible than the nonparametric 
rules \eqref{tukey} and \eqref{kimber}.
Further, as the results of Section 3.2 demonstrate, for asymmetric 
distributions, the lower outliers can be as damaging to parameter 
estimators as the upper outliers are. Therefore, we propose to 
supplement the outlier-labeling rule \eqref{iglewicz1} with 
an analogously designed lower bound:
\begin{equation}
\big[ \widehat{q}_1 - \widehat{K}_1 \cdot 
(\widehat{q}_3 - \widehat{q}_1); \, \widehat{q}_3 + \widehat{K}_2 
\cdot (\widehat{q}_3 - \widehat{q}_1) \big] ~=~ 
\big[ \widehat{F}^{-1} \big( 1-(1-\alpha/2)^{1/n} \big); \, 
\widehat{F}^{-1} \big( (1-\alpha/2)^{1/n} \big) \big],
\label{iglewicz2}
\end{equation}
where 
$\widehat{K}_1 = 
\frac{ \widehat{q}_1 - \widehat{F}^{-1}(1- (1-\alpha/2)^{1/n})}
{\widehat{q}_3-\widehat{q}_1}$ ~and~
$\widehat{K}_2 = 
\frac{ \widehat{F}^{-1}((1-\alpha/2)^{1/n}) - \widehat{q}_3}
{\widehat{q}_3-\widehat{q}_1}$,
and the term 
$F^{-1} \big( 1-(1-\alpha/2)^{1/n} \big)$ is derived by solving 
$\alpha/2 = \mbox{\bf P} \big[ X_{(1)} \leq t \big] = 
1 - [ 1 - F(t)) ]^n$ for $t$.
\end{itemize}

Outcomes of the implementation of the outlier-labeling rules 
\eqref{tukey}-\eqref{iglewicz2} on the 8\% contaminated Weibull data 
(see Tables 3.1 and 3.2) are presented in Table 3.3, where we report 
the proportions of declared outliers in three distinct intervals: 
$[0; \, 4.34 \cdot 10^{-3}]$ ({\em lower\/} outliers),
$(4.34 \cdot 10^{-3}; \, 54.4)$ ({\em non-extreme\/} data), and
$(54.4; \, \infty)$ ({\em upper\/} outliers). Also, since the data 
are simulated from a fully-specified distribution, the true values of 
parameters of $f_0$ are known. Therefore, we can apply \eqref{iglewicz1} 
and \eqref{iglewicz2} with $\theta$ and $\tau$ estimated by MLE, 
log-oQLS, and log-gQLS and compare those decisions with the decision 
based on $\theta_0 = 5, \, \tau_0 = 3/4$ (this will be called an 
`oracle decision').

\begin{center}
{\sc Table 3.3.} Performance of Tukey \eqref{tukey}, Kimber \eqref{kimber}, 
and Banerjee-Iglewicz \eqref{iglewicz1} and \eqref{iglewicz2} 
\\[-1ex]
outlier-labeling rules when applied on the 8\% contaminated Weibull 
data. For \eqref{iglewicz1} and \eqref{iglewicz2}, 
\\[-1ex]
$\alpha = 0.05$ and o2/g2 denote log-oQLS/log-gQLS estimators 
based on $a=1-b=0.05$.

\medskip

\begin{tabular}{|c|c|c|c|c|c|}
\hline
Method & Parameter & Screening & 
\multicolumn{3}{|c|}{Proportion of Outliers} \\[-0.5ex]
\cline{4-6}
 & Values & Interval & Lower & Non-extreme & Upper \\
\hline
\hline
Tukey \eqref{tukey}   & -- -- -- & $[-11.2; \, 20.8]$ & 0/7 & 25/484 & 9/9 \\
Kimber \eqref{kimber} & -- -- -- & $[-2.6; \, 17.4]$ & 0/7 & 38/484 & 9/9 \\
\hline
Banerjee-Iglewicz & $\theta_0 = 5, \, \tau_0 = 3/4$ & $[0; \, 95.2]$ & 
0/7 & 0/484 & 8/9 \\
\eqref{iglewicz1} & $\widehat{\theta}_{\mbox{\tiny MLE}} = 5.66, \, 
 \widehat{\tau}_{\mbox{\tiny MLE}} = 0.49$ & $[0; \, 521.2]$ & 
0/7 & 0/484 & 4/9 \\
 & $\widehat{\theta}_{\mbox{\tiny o2}} = 5.14, \, 
 \widehat{\tau}_{\mbox{\tiny o2}} = 0.66$ & $[0; \, 147.6]$ & 
0/7 & 0/484 & 7/9 \\
 & $\widehat{\theta}_{\mbox{\tiny g2}} = 5.10, \, 
 \widehat{\tau}_{\mbox{\tiny g2}} = 0.66$ & $[0; \, 143.3]$ & 
0/7 & 0/484 & 7/9 \\
\hline
Banerjee-Iglewicz & $\theta_0 = 5, \, \tau_0 = 3/4$ &
$[9.4 \cdot 10^{-6}; \, 106.2]$ & 1/7 & 0/484 & 8/9 \\
\eqref{iglewicz2} & $\widehat{\theta}_{\mbox{\tiny MLE}} = 5.66, \, 
 \widehat{\tau}_{\mbox{\tiny MLE}} = 0.49$ & $[9.2 \cdot 10^{-9}; \, 616.0]$ & 
0/7 & 0/484 & 4/9 \\
 & $\widehat{\theta}_{\mbox{\tiny o2}} = 5.14, \, 
 \widehat{\tau}_{\mbox{\tiny o2}} = 0.66$ & $[1.5 \cdot 10^{-6}; \, 167.1]$ & 
1/7 & 0/484 & 7/9 \\
 & $\widehat{\theta}_{\mbox{\tiny g2}} = 5.10, \, 
 \widehat{\tau}_{\mbox{\tiny g2}} = 0.66$ & $[1.7 \cdot 10^{-6}; \, 162.1]$ & 
1/7 & 0/484 & 7/9 \\
\hline
\end{tabular}
\end{center}

\medskip

Several observations can be made from this exercise. 
First, \eqref{tukey}-\eqref{iglewicz1} work poorly in identifying lower 
outliers, as expected. Second, the nonparametric rules \eqref{tukey} and 
\eqref{kimber} are successful in identifying all upper outliers, but that 
comes at the expense of declaring many outliers in the non-extreme region. 
Third, \eqref{iglewicz1} and \eqref{iglewicz2} handle the non-extreme region 
well. MLE-based decisions are inferior to those based on log-oQLS and log-gQLS. 
For Banerjee-Iglewicz \eqref{iglewicz2}, performances of the robust log-oQLS 
and log-gQLS rules get reasonably close to that of the oracle rule. 

\begin{center}
{\sc Table 3.4.} Performance of Tukey \eqref{tukey}, 
Kimber \eqref{kimber}, and Banerjee-Iglewicz \eqref{iglewicz1} 
and \eqref{iglewicz2} 
\\[-1ex]
when applied on 8\% contaminated Weibull and lognormal data. 
$M=10^4$ samples of size $n=500$. 
\\[-1ex]
For \eqref{iglewicz1} and \eqref{iglewicz2}, $\alpha = 0.05$ 
and o2/g2 denote log-oQLS/log-gQLS based on $a=1-b=0.05$.

\medskip

\begin{tabular}{|c|c|c|c|c|c|}
\hline
Contaminated & Method & Parameter & 
\multicolumn{3}{|c|}{Proportion of Declared Outliers} \\[-0.5ex]
\cline{4-6}
Distribution & & Estimation & Lower & Non-extreme & Upper \\
\hline
\hline
{\em Weibull\/} ($f_{0.08}$) with & Tukey \eqref{tukey} & -- -- -- & 
0 & 0.07 & 1.00 \\
{\footnotesize $f_0$: We$\, (\theta_0 = 5, \tau_0 = 3/4)$} &
Kimber \eqref{kimber} & -- -- -- & 0 & 0.10 & 1.00 \\
\cline{2-6}
{\footnotesize $f_1$: We$\, (\theta_1 = 5, \tau_1 = 1/4)$} &
Banerjee-Iglewicz & Oracle & 0 & 0 & 0.81 \\
 & \eqref{iglewicz1} & MLE & 0 & 0 & 0.44 \\
 & & o2 & 0 & 0 & 0.67 \\
 & & g2 & 0 & 0 & 0.69 \\
 \cline{2-6}
 & Banerjee-Iglewicz & Oracle & 0.37 & 0 & 0.77 \\
 & \eqref{iglewicz2} & MLE & 0.12 & 0 & 0.40 \\
 & & o2 & 0.26 & 0 & 0.63 \\
 & & g2 & 0.27 & 0 & 0.65 \\
\hline
{\em Lognormal\/} ($f_{0.08}$) with & Tukey \eqref{tukey} & -- -- -- & 
0 & 0.08 & 1.00 \\
{\footnotesize $f_0$: LN$\, (\mu_0 = 0, \sigma_0 = 1)$} &
Kimber \eqref{kimber} & -- -- -- & 0 & 0.10 & 1.00 \\
\cline{2-6}
{\footnotesize $f_1$: LN$\, (\mu_1 = 2, \sigma_1 = 2)$} &
Banerjee-Iglewicz & Oracle & 0 & 0 & 0.66 \\
 & \eqref{iglewicz1} & MLE & 0 & 0 & 0.29 \\
 & & o2 & 0 & 0 & 0.45 \\
 & & g2 & 0 & 0 & 0.43 \\
\cline{2-6}
 & Banerjee-Iglewicz & Oracle & 0.24 & 0 & 0.58 \\
 & \eqref{iglewicz2} & MLE & 0.06 & 0 & 0.23 \\
 & & o2 & 0.14 & 0 & 0.38 \\
 & & g2 & 0.13 & 0 & 0.37 \\
\hline
\multicolumn{6}{l}{\footnotesize {\sc Note}: ~For Weibull, 
329 (out of $10^4$) samples contained no outliers; 
they were excluded from computations.} 
\\[-1ex]
\multicolumn{6}{l}{\footnotesize For lognormal, 
7574 (out of $10^4$) samples contained no outliers; 
they were excluded from computations.} \\
\end{tabular}
\end{center}

\medskip

Of course, these observations are based on only one sample and thus can 
be coincidental. To make the analysis more reliable, we repeated the 
exercise on $M=10^4$ samples of size $n=500$. We also broadened the 
study by considering two distributions: 8\% contaminated Weibull and 
8\% contaminated lognormal. The results are summarized in Table 3.4. 
Overall, the conclusions stated above do not change. However, among 
$M=10^4$ samples there were such that generated either no lower or 
upper outliers, i.e., ``clean'' in at least one tail. For Weibull, 
there were 329 such samples, and for lognormal, 7574 samples. They 
were excluded from further computations. Hence, the reported outlier 
detection rates are valid for the samples that actually contained 
outliers in both tails.

\subsection{Goodness of Fit}

The next study evaluates the power properties of the goodness-of-fit 
test \eqref{gof2} under several alternatives. The results are presented 
in Table 3.5. The following conclusions emerge from the table.

First of all, the test is correctly calibrated because for both 
distributions and for all choices of $(a, b)$ and all sample sizes, 
it rejects the null hypothesis (i.e., when $f_{\varepsilon} = f_0$) 
$5\%$ of the time, which agrees with its level of significance 
($\alpha = 0.05$).
Secondly, under the contaminated data scenarios (i.e., when  
$f_{\varepsilon} \ne f_0$), the rejection rates increase and
approach 1 as $n \rightarrow 1000$. This happens quite quickly 
for $f_{0.08}$.
Thirdly, performance of the test in the $f_{0.03}$ case of
Weibull is fairly steady, with the rejection rates fluctuating 
around 0.20. And for the lognormal distribution, the test starts 
slowly at $n=100$, but then its performance improves rapidly.
Finally, for both distributions and under all $\varepsilon > 0$ 
and $n$, higher values of $a = 1-b$ lead to slightly higher 
rejection rates until they reach 1.
 
\smallskip

\begin{center}
{\sc Table 3.5.} Proportions of rejections of $H_0$ by the 
goodness-of-fit test \eqref{gof2} at $\alpha = 0.05$ for
\\[-1ex]
Weibull and lognormal distributions under $H_0$ and $H_A$, 
and varying $n$. For log-gQLS (`g') 
\\[-1ex]
estimators, $(a, b)$ are:
$(0.02, 0.98)$ for g1, 
$(0.05, 0.95)$ for g2, 
and $(0.10, 0.90)$ for g3.

\medskip

\begin{tabular}{|c|c|ccc|ccc|ccc|}
\hline
Assumed & Parameter & \multicolumn{9}{|c|}{Data Generated by
Contaminated Distribution $f_{\varepsilon}$} \\[-0.5ex]
Distribution & Estimation & $f_0$ & $f_{0.03}$ & $f_{0.08}$ & 
$f_0$ & $f_{0.03}$ & $f_{0.08}$ & $f_0$ & $f_{0.03}$ & $f_{0.08}$ \\
\cline{3-11}
 & & \multicolumn{3}{|c|}{$n=100$} & \multicolumn{3}{|c|}{$n=500$} &
\multicolumn{3}{|c|}{$n=1000$} \\
\hline
\hline
{\em Weibull\/} ($f_0$) & 
   g1 & 0.05 & 0.18 & 0.36 & 0.05 & 0.15 & 0.71 & 0.05 & 0.19 & 0.91 \\
 & g2 & 0.05 & 0.20 & 0.42 & 0.05 & 0.18 & 0.76 & 0.05 & 0.22 & 0.94 \\
 & g3 & 0.05 & 0.21 & 0.45 & 0.05 & 0.19 & 0.78 & 0.05 & 0.24 & 0.95 \\
\hline
{\em Lognormal\/} ($f_0$) &  
   g1 & 0.05 & 0.10 & 0.22 & 0.05 & 0.32 & 0.91 & 0.05 & 0.56 & 1 \\
 & g2 & 0.05 & 0.12 & 0.29 & 0.05 & 0.35 & 0.94 & 0.05 & 0.60 & 1 \\
 & g3 & 0.05 & 0.13 & 0.34 & 0.05 & 0.37 & 0.95 & 0.05 & 0.62 & 1 \\
\hline
\end{tabular}
\end{center}

\medskip

\section{Real Data Examples}

To illustrate how our proposed estimators, outlier-labeling rules, 
and the goodness-of-fit test work on real data, we use two data sets: 
the daily stock returns of Google for 2020--2023 (Section 4.1) and 
the normalized losses of the most damaging hurricanes in the U.S. 
from 1900 to 2022 (Section 4.2).

\subsection{Google Returns}

We start by analyzing the {\em daily stock returns\/} of Alphabet Inc., 
the parent company of {\em Google\/}, for the period from January 2, 2020, 
to December 29, 2023. The stock prices are available at the 
{\em Yahoo!\/}{\em Finance\/} website 
{\tt ~https://finance.yahoo.com/quote/GOOG/}. 
The daily returns are computed by taking the ratio of the closing 
over the opening price. Below are summary statistics for the 
log-transformed data.

\medskip

\begin{center}
\begin{tabular}{c|ccccc|cc}
Sample Size & \multicolumn{5}{|c|}{Five-Number Summary} &
\multicolumn{2}{|c}{Moments} \\[-0.5ex]
$n$ & $min$ & $q_1$ & $q_2$ & $q_3$ & $max$ & $mean$ & $std. dev.$ \\
\hline
$1006$ & -0.0574 & -0.0084 & $0.0014$ & $0.0109$ & $0.0644$ & 
$0.0010$ & $0.0157$ \\
\end{tabular}
\end{center}

\medskip

Although not presented here, the histogram of the log-data revealed 
that a symmetric and (roughly) bell-shaped distribution might be a 
suitable candidate for the dataset at hand. Therefore, log-Cauchy, 
log-Laplace, log-Logistic, and lognormal distributions will be 
fitted to the daily returns using log-gQLS. 
To determine what levels of $a$ and $b$ are appropriate in this case, 
we employed the nonparametric outlier-labeling rules of Section 3.3.
For Tukey \eqref{tukey}, the screening interval is [-0.037; 0.039], 
which implies that there are 12 lower and 11 upper outliers (out 
of $n = 1006$ observations). This is equivalent to $1.2\%$ lower 
and $1.1\%$ upper contamination. 
For Kimber \eqref{kimber}, the screening interval is [-0.023; 0.025], 
which declares 64 lower and 55 upper outliers. Or equivalently, 
$6.4\%$ lower and $5.5\%$ upper contamination.
Given that the two methods disagree on the levels of contamination, 
we select $a = 1-b = 0.02, \, 0.05, \, 0.10$. This mixed selection 
of $a$ and $b$ will show us if it has any material effect on further 
analysis. Note that all these choices are safe according to 
\eqref{tukey}, but they offer varying levels of safety according 
to \eqref{kimber}.

Next, model fits are formally validated using the goodness-of-fit 
test \eqref{gof2}, for which the quantile levels (for model 
validation) are: ~$p_1^{\mbox{\tiny out}} = 0.01, \, 
p_2^{\mbox{\tiny out}} = 0.03, \ldots, \, 
p_{49}^{\mbox{\tiny out}} = 0.97, \, 
p_{50}^{\mbox{\tiny out}} = 0.99$.
The findings of this analysis are summarized in Table 4.1.
It is clear from the table that the log-Logistic distribution provides 
the best fit for this dataset, as its $p$-values are significantly 
greater than $0.10$ for all three log-gQLS estimators. The log-Laplace 
model is borderline acceptable, with its $p$-values ranging between 
0.06 and 0.11. The log-Cauchy and lognormal distributions are strongly 
rejected by the test \eqref{gof2}. Also, the choice of proportions $a$ 
and $b$ has negligible effect on the goodness-of-fit of all models.

\newpage

\begin{center}
{\sc Table 4.1.} Parameter estimates and goodness-of-fit 
statistics for various log-location-scale
\\[-1ex]
families fitted to the daily log-returns of Google stock.
Proportions $(a, b)$ are: $(0.02, 0.98)$
\\[-1ex]
for g1, $(0.05, 0.95)$ for g2, and 
$(0.10, 0.90)$ for g3. Also, $k=25$ and $m=50$.

\medskip

\begin{tabular}{|c|c|cc|c|}
\hline
Assumed & Method of & \multicolumn{2}{|c|}{Parameter Estimates} &  
Goodness of Fit \\[-0.5ex]
Distribution & Estimation &  ~~~~ $\widehat{\mu}$ ~~ & 
~~~~ $\widehat{\sigma}$ ~~ & 
$W_{\mbox{\tiny out}}$ {\footnotesize ($p$-value)} \\
\hline
\hline
{\em Log-Cauchy\/} & g1 & 0.0014 & 0.0091 & 83.77 {\footnotesize (0.01)} \\
& g2 & 0.0014 & 0.0092 & 82.43 {\footnotesize (0.01)} \\
& g3 & 0.0014 & 0.0092 & 81.89 {\footnotesize (0.01)} \\
\hline
{\em Log-Laplace\/} & g1 & 0.0014 & 0.0121 & 65.93 {\footnotesize (0.06)} \\
& g2 & 0.0014 & 0.0124 & 63.54 {\footnotesize (0.08)} \\
& g3 & 0.0014 & 0.0127 & 62.01 {\footnotesize (0.11)} \\
\hline
{\em Log-Logistic\/} & g1 & 0.0012 & 0.0087 & 
{\bf 40.26} {\footnotesize ({\bf 0.78})} \\
& g2 & 0.0012 & 0.0087 & {\bf 39.73} {\footnotesize ({\bf 0.80})} \\
& g3 & 0.0012 & 0.0087 & {\bf 39.92} {\footnotesize ({\bf 0.79})} \\
\hline
{\em Lognormal\/} & g1 & 0.0010 & 0.0152 & 70.64 {\footnotesize (0.02)} \\
& g2 & 0.0011 & 0.0150 & 73.73 {\footnotesize (0.01)} \\
& g3 & 0.0010 & 0.0147 & 82.11 {\footnotesize (0.00)} \\
\hline
\end{tabular}
\end{center}

\medskip

Further, as the findings of Section 3.3 suggest, the nonparametric 
outlier-labeling rules have a tendency to declare too many outliers.
Some of those declarations are likely to be false positives. 
Thus, equipped with the conclusions of Table 4.1, we will identify 
outliers using the model-based approach of Banerjee-Iglewicz 
\eqref{iglewicz2}. Since the log-Logistic and log-Laplace models 
offer the best fits, they will be used to identify outliers. 
The lognormal distribution will also be included in this analysis 
because it plays a prominent role in many financial risk management 
applications. To better understand the importance of parameter 
estimation methods, these models will be fitted using MLE, log-oQLS 
and log-gQLS estimators. The results of this analysis are summarized 
in Table 4.2.

As is evident from Tables 4.1 and 4.2, the log-Laplace and 
log-Logistic distributions fit this dataset well and therefore
are able to accommodate the numerous nonparametrically identified 
outliers, 64 lower and 55 upper according to \eqref{iglewicz2}. 
For these models, all estimation methods (including MLE) lead 
to practically the same inference and decisions. The lognormal 
distribution, on the other hand, does not fit this dataset well.
This seems to be due to a single {\em upper\/} outlier, which is 
successfully identified by the robust log-oQLS and log-gQLS 
estimators, but missed by MLE.

\newpage

\begin{center}
{\sc Table 4.2.} Model-based outlier labeling by Banerjee-Iglewicz 
\eqref{iglewicz2} at $\alpha = 0.05$ for the daily 
\\[-1ex]
log-returns of Google stock. For log-oQLS/log-gQLS (`o'/`g'), 
proportions $(a, b)$ are: $(0.02, 0.98)$ 
\\[-1ex]
for o1/g1, $(0.05, 0.95)$ for o2/g2, 
and $(0.10, 0.90)$ for o3/g3.
In all cases, $k=25$.

\medskip

\begin{tabular}{|c|c|c|c|c|}
\hline
Assumed & Parameter & Screening & 
\multicolumn{2}{|c|}{Proportion of Declared Outliers} \\[-0.5ex]
\cline{4-5}
Distribution & Estimation & Interval & $\quad$ ~Lower~ $\quad$ & 
$\quad$ ~Upper~ $\quad$ \\
\hline
\hline
{\em Log-Laplace\/} & MLE & [-0.1172; 0.1199] & 0/1006 & 0/1006 \\
 & o1 & [-0.1105; 0.1124] & 0/1006 & 0/1006 \\
 & o2 & [-0.1175; 0.1197] & 0/1006 & 0/1006 \\
 & o3 & [-0.1244; 0.1267] & 0/1006 & 0/1006 \\
 & g1 & [-0.1185; 0.1212] & 0/1006 & 0/1006 \\
 & g2 & [-0.1212; 0.1239] & 0/1006 & 0/1006 \\
 & g3 & [-0.1245; 0.1272] & 0/1006 & 0/1006 \\
\hline
{\em Log-Logistic\/} & MLE & [-0.0906; 0.0929] & 0/1006 & 0/1006 \\
 & o1 & [-0.0904; 0.0924] & 0/1006 & 0/1006 \\
 & o2 & [-0.0911; 0.0933] & 0/1006 & 0/1006 \\
 & o3 & [-0.0909; 0.0932] & 0/1006 & 0/1006 \\
 & g1 & [-0.0905; 0.0928] & 0/1006 & 0/1006 \\
 & g2 & [-0.0910; 0.0934] & 0/1006 & 0/1006 \\
 & g3 & [-0.0908; 0.0932] & 0/1006 & 0/1006 \\
\hline
{\em Lognormal\/} & MLE & [-0.0626; 0.0646] & 0/1006 & 0/1006 \\
 & o1 & [-0.0610; 0.0629] & 0/1006 & 1/1006 \\
 & o2 & [-0.0593; 0.0615] & 0/1006 & 1/1006 \\
 & o3 & [-0.0576; 0.0599] & 0/1006 & 1/1006 \\
 & g1 & [-0.0607; 0.0627] & 0/1006 & 1/1006 \\
 & g2 & [-0.0599; 0.0620] & 0/1006 & 1/1006 \\
 & g3 & [-0.0584; 0.0605] & 0/1006 & 1/1006 \\
\hline
\end{tabular}
\end{center}

\bigskip

Finally, the single upper outlier, as identified by the robustly 
fit lognormal model, occurred on November 30, 2022. On that day 
the opening and closing prices were 95.12 and 101.45, respectively, 
resulting in the log-return of 0.0644. This outlier-level increase 
did coincide with a stocks rally on Wall Street after the head 
of the Federal Reserve said that the central bank plans to make 
some favorable adjustments to the interest rates. Major indexes 
experienced substantial but not extraordinary gains. For example, 
the S\&P 500 rose 122.48 points, or 3.1\%, to 4,080.11. And 
the Dow Jones Industrial Average rose 737.24 points, or 2.2\%, 
to 34,589.77.

\subsection{Hurricane Damages}

For the second illustration, we use the {\em normalized\/} losses from 
the most damaging hurricanes in the United States from 1900 to 2022, 
as recorded by \citet[][]{metal25}; see their data availability 
statement for the GitHub repository address where the data sets 
are stored. The damages were adjusted to 2022 dollars by inflation, 
wealth, and coastal population changes using two types of normalization: 
the original one proposed by \citet[][]{pl98}, abbreviated PL22, and 
an alternative normalization of \citet[][]{cl01}, abbreviated CL22.
The PL22-normalized damages are computed as follows:
\[
D_{\mbox{\tiny PL22}} ~=~ 
D_y \times I_y \times \mbox{RWPC}_y \times P_{2022/y},
\]
where $D_y$ is the reported damage in landfall-year U.S. dollars, 
$I_y$ is an inflation adjustment, $\mbox{RWPC}_y$ is the real 
wealth per capita adjustment, and $P_{2022/y}$ is the county 
population adjustment. And the CL22 normalization is defined
by the following formula:
\[
D_{\mbox{\tiny CL22}} ~=~ 
D_y \times I_y \times \mbox{RWHU}_y \times \mbox{HU}_{2022/y},
\]
where $D_y$ and $I_y$ are defined as before, $\mbox{RWHU}_y$ 
is the real wealth per housing unit adjustment, and 
$\mbox{HU}_{2022/y}$ is the county housing unit adjustment.
See the above-cited references for more details.

Further, it was concluded by \citet[][]{metal25} that ``these 
normalized losses showed no obvious trends over time''. To check 
this conclusion, we split the data sets in two periods, 1900-1999
and 2000-2022, apply our model fitting and validation methodologies
on the 1900-1999 period and then use the best-fitting distributions 
to predict tail probabilities for the 2000-2022 period. (This choice 
of the time periods could be labeled as the 20th century versus 
the 21st century.) If there is indeed no trend over time, the 
predictions based on well-constructed models should be close to 
the actually observed tail probabilities in the 21st century. Below 
are summary statistics for the PL22- and CL22-normalized losses 
over the period 1900-1999 ($n=32$ hurricanes; damages measured in 
billions of 2022 U.S. dollars).

\medskip

\begin{center}
\begin{tabular}{c|ccccc|cc}
Data & \multicolumn{5}{|c|}{Five-Number Summary} &
\multicolumn{2}{|c}{Moments} \\[-0.5ex]
Normalization & $min$ & $q_1$ & $q_2$ & $q_3$ & $max$ & $mean$ & $std. dev.$ \\
\hline
PL22 & 8.43 & 19.16 & 42.22 & 63.86 & 206.97 & 54.01 & 48.43 \\
CL22 & 6.88 & 20.70 & 36.93 & 60.50 & 178.48 & 51.69 & 43.35 \\
\end{tabular}
\end{center}

\medskip

Preliminary diagnostics such as histograms of damages and log-damages 
showed that an approximately log-symmetric distribution might be 
appropriate for both data sets. Therefore, log-Cauchy, log-Gumbel 
(Weibull), log-Laplace, log-Logistic, and lognormal distributions 
will be fitted to these data using log-gQLS. Also, the nonparametric 
outlier-labeling tests yield between 3 and 5 upper outliers out of 
$n=32$ observations (equivalently, between 9.4\% and 15.6\% upper 
contamination). Specifically, Tukey \eqref{tukey} outlier-screening 
intervals are [-47.9; 130.9] (for PL22) and [-39.0; 120.2] (for CL22), 
and Kimber \eqref{kimber} intervals are [-15.4; 96.3] (for PL22) and 
[-3.7; 95.9] (for CL22). Since $n=32$ is not large, the proportions
$a$ and $b$ change quite a bit due to inclusion or exclusion of a 
single sample point. Therefore, we select $(a, b) = (0.10, 0.90)$, 
which is a more conservative choice for the lower tail but in the 
range of the \eqref{tukey} and \eqref{kimber} recommendations for
the upper tail.

Furthermore, in Table 4.3, estimates of the parameters and results of 
the goodness-of-fit test \eqref{gof2} are provided for the selected
distributions. For model estimation, we choose $k=15$ and for model 
validation, $m=15$ with the quantile levels 
$p_1^{\mbox{\tiny out}} = 0.050, \, 
p_2^{\mbox{\tiny out}} = 0.114, \ldots, \, 
p_{14}^{\mbox{\tiny out}} = 0.886, \, 
p_{15}^{\mbox{\tiny out}} = 0.950$.
Clearly (although perhaps surprisingly), the results in the table 
suggest that all log-gQLS-estimated models fit both data sets well 
with their respective $p$-values substantially exceeding 0.10. 
Therefore, as one would expect, Banerjee-Iglewicz \eqref{iglewicz2} 
identifies zero model-specific outliers in each case. Thus according 
to these log-location-scale distributions, both data sets are ``clean''.

\begin{center}
{\sc Table 4.3.} Parameter estimates and goodness-of-fit 
statistics for selected log-location-scale
\\[-1ex]
families fitted to the normalized damages via log-gQLS 
with $(a, b) = (0.10, 0.90)$, $k=15$.

\medskip

\begin{tabular}{|c|c|cc|c|}
\hline
Data & Assumed & 
\multicolumn{2}{|c|}{Parameter Estimates} &  Goodness of Fit \\[-0.5ex]
Normalization & Distribution &  
~~~~ $\widehat{\mu}$ ~~ & ~~~~ $\widehat{\sigma}$ ~~ & 
$W_{\mbox{\tiny out}}$ {\footnotesize ($p$-value)} \\
\hline
\hline
PL22 & {\em Log-Cauchy\/} & ~24.406 & ~0.582 & 10.91 {\footnotesize (0.68)} \\
 & {\em Log-Gumbel\/}     & ~23.982 & ~0.769 & 12.41 {\footnotesize (0.59)} \\
 & {\em Log-Laplace\/}    & ~24.408 & ~0.770 & 15.20 {\footnotesize (0.44)} \\
 & {\em Log-Logistic\/}   & ~24.352 & ~0.514 & 12.90 {\footnotesize (0.55)} \\
 & {\em Lognormal\/}      & ~24.354 & ~0.863 & 12.91 {\footnotesize (0.55)} \\
\hline
CL22 & {\em Log-Cauchy\/} & ~24.293 & ~0.617 & 15.55 {\footnotesize (0.47)} \\
 & {\em Log-Gumbel\/}     & ~23.963 & ~0.796 & 15.23 {\footnotesize (0.41)} \\
 & {\em Log-Laplace\/}    & ~24.270 & ~0.812 & 20.16 {\footnotesize (0.22)} \\
 & {\em Log-Logistic\/}   & ~24.337 & ~0.544 & 13.83 {\footnotesize (0.49)} \\
 & {\em Lognormal\/}      & ~24.353 & ~0.909 & 13.36 {\footnotesize (0.52)} \\
\hline
\end{tabular}
\end{center}

\bigskip

The final step in this analysis is to examine the ``no trends'' conclusion.
Using 1900-1999 data and the fitted models, we predict the probability 
that the normalized damage of a hurricane happening from 2000-2022 exceeds 
a certain (high) threshold. For context, recall that during the first two 
decades of this century there were some devastating hurricanes in the 
United States.
\begin{itemize}
  \item[$*$] {\em Hurricane Katrina\/} (2005). It was category 5 at peak 
and 3 at landfall in Louisiana, with reported damages of $125$ billion 
(2005 USD) and $1392$ deaths (and 652 missing). Katrina devastated New 
Orleans due to levee failures. It is considered one of the costliest 
natural disasters in U.S. history. After normalization, its damages are 
$D_{\mbox{\tiny PL22}} = 226.21$ and $D_{\mbox{\tiny CL22}} = 234.11$
billion.

  \item[$*$] {\em Hurricane Harvey\/} (2017). It was category 4 at 
landfall in Texas, with reported damages of $125$ billion (2017 USD) 
and 107 deaths. Harvey brought very heavy rainfall (up to 60 inches in 
some areas) and caused massive flooding in Houston. After normalization, 
its damages are $D_{\mbox{\tiny PL22}} = 164.70$ and 
$D_{\mbox{\tiny CL22}} = 161.12$ billion.
\end{itemize}

In view of the above discussion, we consider the following normalized
thresholds: 50, 100, 150, 200 billion. We also compare the model-based 
predictions with the simple empirical predictions. The results are 
summarized in Table 4.4. As one can see in the table, all methods 
work well for $t=50$ but for higher $t$ their performances diverge. 
The empirical approach, the lognormal and log-logistic models 
underestimate the probabilities of exceeding $t = 100, ~150, ~200$. 
The log-Cauchy tends to overestimate the probabilities of extreme 
thresholds (see its predictions for $t = 150$ and $200$). The 
log-Gumbel and log-Laplace distributions -- fitted via log-gQLS with 
$a=0.10$, $b=0.90$, $k=15$ -- emerge as the most effective models 
for this task among the six approaches considered here. Also, for 
practically all the entries in Table 4.4, the actual probabilities 
fall within one standard deviation of the predicted point estimate. 
For example: 
$0.159 \pm 0.055$ (for Log-Gumbel, PL22, $t=100$) covers the actual 
probability of 0.182, and $0.031 \pm 0.031$ (for Empirical, PL22, 
$t=200$) includes 0.045. 
Overall, the ``no trends'' conclusion of \citet[][]{metal25} seems 
reasonable but more thorough investigations may change it.

\begin{center}
{\sc Table 4.4.} Empirical and parametric predictions of 
$\mbox{\bf P} \big[ D_{\mbox{\tiny PL22}} > t \big]$ 
and
$\mbox{\bf P} \big[ D_{\mbox{\tiny CL22}} > t \big]$
for 2000-2022.
\\[-1ex]
The loss models are fitted via log-gQLS with 
$(a, b) = (0.10, 0.90)$, $k=15$, and 1900-1999 data.

\medskip

\begin{tabular}{|c|cccc|cccc|}
\hline
Method of & \multicolumn{4}{|c|}{PL22 Normalization} &  
\multicolumn{4}{|c|}{CL22 Normalization} \\[-0.5ex]
\cline{2-9}
Prediction & $t=50$ & $t=100$ & $t=150$ & $t=200$ & 
$t=50$ & $t=100$ & $t=150$ & $t=200$ \\
\hline
\hline
Empirical & 0.406 & 0.156 & 0.063 & 0.031 & 
0.438 & 0.156 & 0.031 & 0.000 \\
{\em Log-Cauchy\/}   & 0.381 & 0.179 & 0.132 & 0.110 & 
0.339 & 0.171 & 0.129 & 0.109 \\
{\em Log-Gumbel\/}   & 0.348 & 0.159 & 0.097 & 0.068 & 
0.349 & 0.165 & 0.102 & 0.073 \\
{\em Log-Laplace\/}  & 0.372 & 0.151 & 0.089 & 0.062 & 
0.319 & 0.136 & 0.082 & 0.058 \\
{\em Log-Logistic\/} & 0.365 & 0.130 & 0.064 & 0.037 & 
0.366 & 0.139 & 0.071 & 0.043 \\
{\em Lognormal\/}    & 0.372 & 0.129 & 0.055 & 0.027 & 
0.378 & 0.141 & 0.064 & 0.033 \\
\hline
{\sc Actual} & 0.409 & 0.182 & 0.091 & 0.045 & 
0.318 & 0.182 & 0.091 & 0.045 \\
\hline
\end{tabular}
\end{center}

\section{Concluding Remarks}

In this paper, four {\em quantile\/} and {\em log-quantile least 
squares\/} estimators for the parameters of log-location-scale 
loss models have been introduced: ordinary (oQLS and log-oQLS) and 
generalized (gQLS and log-gQLS). While all estimators are equally 
robust, the ones based on log-quantiles are more efficient than 
the corresponding oQLS and gQLS estimators. Moreover, log-oQLS and 
log-gQLS are computationally inexpensive -- they can be computed 
on a basic laptop in 2-3 minutes for samples containing a 
{\em billion\/} observations.
In terms of the estimation accuracy, both ``log'' methods are 
consistent but log-gQLS exhibits substantially smaller sampling 
errors than log-oQLS and often reaches 90\% or higher relative 
efficiency when compared to the maximum likelihood estimators 
which are theoretically optimal. Nonetheless, the log-oQLS 
estimator (being practically a simpler version of log-gQLS) may 
be easier to extend to more general loss models. 
Further, the advantages of log-oQLS and log-gQLS over MLE become 
obvious when the underlying distributional assumptions are violated 
by the presence of outliers. These results and properties have been 
established using theoretical derivations (when sample size $n$ is 
large) and augmented by using simulations. Furthermore, we have 
applied these estimators to several data-screening rules and 
verified their effectiveness in identifying outliers. We have 
observed that the success rates of log-oQLS- and log-gQLS-based 
rules are significantly higher than those based on MLE and slightly 
below the rates of the oracle rule (which mimics the scenario when 
the assumed underlying model is completely known). Similar conclusions 
have been reached when these estimators were applied to a newly-designed 
goodness-of-fit test. That is, when sample size exceeds 100 observations, 
the test has sufficient power to detect and reject even mildly contaminated 
distributions. Of course, its power improves further against moderate 
or severe contamination levels.
Finally, additional illustrations involving the proposed estimators, 
outlier-labeling rules, and the goodness-of-fit test have been provided 
using the daily stock returns of Alphabet Inc. (Google) over the years 
2020-2023 and the normalized losses of the most damaging hurricanes 
in the U.S. from 1900-2022.

\section*{Acknowledgments}

The authors are very appreciative of valuable suggestions, technical 
queries, and insightful comments provided by two anonymous referees, 
which helped to substantially improve the paper.


\baselineskip 5mm

\begin{thebibliography}{99}

\bibitem[Adjieteh and Brazauskas(2025)]{ab25}
Adjieteh, M. and Brazauskas, V. (2025).
Quantile least squares: A flexible approach for robust 
estimation and validation of location-scale families.
{\em Statistics and Computing\/}, 
{\bf 35}, article 106, 1--21.

\bibitem[Arnold(2014)]{a14}
Arnold, B.C. (2014). {\em Pareto Distributions\/}.
2nd edition, Chapman \& Hall.

\bibitem[Banerjee and Iglewicz(2007)]{bi07}
Banerjee, S. and Iglewicz, B. (2007).
A simple univariate outlier identification procedure 
designed for large samples.
{\em Communications in Statistics -- Simulation and 
Computation\/}, {\bf 36}(2), 249--263.
\bibitem[Basu {\em et al\/}.(2011)]{bsp11}

Basu, A., Shioya, H., and Park, C. (2011). 
{\em Statistical Inference: The Minimum Distance 
Approach\/}. Chapman \& Hall.

\bibitem[Bates and Watts(1988)]{bw88}
Bates, D. and Watts, D. (1988). 
{\em Nonlinear Regression Analysis and Its Applications\/}. 
Wiley.

\bibitem[Bernard {\em et al\/}.(2024)]{bpv24}
Bernard, C., Pesenti, S.M., and Vanduffel, S. (2024).
Robust distortion risk measures.
{\em Mathematical Finance\/},
{\bf 34}(3), 774--818.

\bibitem[Blanchet {\em et al\/}.(2019)]{blty19}
Blanchet, J., Lam, H., Tang, Q., and Yuan, Z. (2019).
Robust actuarial risk analysis.
{\em North American Actuarial Journal\/},
{\bf 23}(1), 33--63.

\bibitem[Brazauskas {\em et al\/}.(2009)]{bjz09}
Brazauskas, V., Jones, B., and Zitikis, R. (2009).
Robust fitting of claim severity distributions and
the method of trimmed moments.
{\em Journal of Statistical Planning and Inference\/},
{\bf 139}(6), 2028--2043.

\bibitem[Brazauskas and Serfling(2000)]{bs00}
Brazauskas, V. and Serfling, R. (2000). 
Robust and efficient estimation of the tail index of a 
single-parameter Pareto distribution (with discussion). 
{\em North American Actuarial Journal\/}, {\bf 4}(4), 12--27. 
Discussion: {\bf 5}(3), 123--126. Reply: {\bf 5}(3), 126--128.

\bibitem[Chen and Wang(2025)]{cw25}
Chen, Y. and Wang, R. (2025). 
Infinite-mean models in risk management: 
Discussions and recent advances.
{\em Risk Sciences\/},
{\bf 1}, article 100003, 1--13.

\bibitem[Collins and Lowe(2001)]{cl01}
Collins, D. and Lowe, S.P. (2001). 
A macro validation dataset for US hurricane models.
{\em Forum\/} (winter edition), 
Casualty Actuarial Society, 217--251.

\bibitem[Dornheim and Brazauskas(2007, 2011, 2014)]{db07}
Dornheim, H. and Brazauskas, V. (2007).
Robust and efficient methods for credibility when claims 
are approximately gamma-distributed.
{\it North American Actuarial Journal\/},
{\bf 11}(3), 138--158.

\bibitem[Dornheim and Brazauskas(2011)]{db11}
Dornheim, H. and Brazauskas, V. (2011).
Robust-efficient credibility models with heavy-tailed 
claims: A mixed linear models perspective.
{\em Insurance: Mathematics and Economics\/}, 
{\bf 48}(1), 72--84.

\bibitem[Dornheim and Brazauskas(2014)]{db14}
Dornheim, H. and Brazauskas, V. (2014).
Case studies using credibility and corrected adaptively 
truncated likelihood methods. 
{\em Variance\/}, {\bf 7}(2), 168--192.

\bibitem[Fung(2022, 2025)]{f22}
Fung, T.C. (2022).
Maximum weighted likelihood estimator for robust 
heavy-tail modelling of finite mixture models.
{\em Insurance: Mathematics and Economics\/},
{\bf 107}, 180--198.

\bibitem[Fung(2025)]{f25}
Fung, T.C. (2025).
Robust estimation and diagnostic of generalized linear 
model for insurance losses: a weighted likelihood 
approach.
{\em Metrika\/}, {\bf 88}(2), 149--182.

\bibitem[Gisler and Reinhard(1993)]{gr93}
Gisler, A. and Reinhard, P. (1993).
Robust credibility.
{\em ASTIN Bulletin\/},
{\bf 23}(1), 118--143.

\bibitem[Hampel {\em et al\/}.(1986)]{hrrs86}
Hampel, F.R., Ronchetti, E.M., Rousseeuw, P.J., 
and Stahel, W.A. (1986).
{\em Robust Statistics: The Approach Based on 
Influence Functions\/}. Wiley.

\bibitem[Huber and Ronchetti(2009)]{hr09}
Huber, P.J. and Ronchetti, E.M. (2009).
{\em Robust Statistics\/}, 2nd edition.
Wiley.

\bibitem[Kim and Jeon(2013)]{kj13} 
Kim, J.H.T. and Jeon, Y. (2013).
Credibility theory based on trimming.
{\it Insurance: Mathematics and Economics\/},
{\bf 53}(1), 36--47.

\bibitem[Kimber(1990)]{k90}
Kimber, A.C. (1990).
Exploratory data analysis for possibly censored data 
from skewed distributions.
{\em Applied Statistics\/}, 
{\bf 39}(1), 21--30.

\bibitem[Klugman {\em et al\/}.(2012)]{kpw12} 
Klugman, S.A., Panjer, H.H., and Willmot, G.E. (2012).
{\em Loss Models: From Data to Decisions\/}.
4th edition, Wiley.

\bibitem[K{\"{u}}nsch(1992)]{k92}   
K{\"{u}}nsch, H.R. (1992).
Robust methods for credibility.
{\em ASTIN Bulletin\/},
{\bf 22}(1), 33--49.

\bibitem[Levenberg(1944)]{l44}
Levenberg, K. (1944).
A method for the solution of certain nonlinear problems 
in least squares.
{\em Quarterly of Applied Mathematics\/}, 
{\bf 2}, 164--168.

\bibitem[Liu and Mao(2022)]{lm22}
Liu, H. and Mao, T. (2022).
Distributionally robust reinsurance with Value-at-Risk 
and Conditional Value-at-Risk.
{\em Insurance: Mathematics and Economics\/},
{\bf 107}, 393--417.

\bibitem[Maronna {\em et al\/}.(2006)]{mmy06}
Maronna, R.A., Martin, R.D., and Yohai, V.J. (2006).
{\em Robust Statistics: Theory and Methods\/}.
Wiley.

\bibitem[Marquardt(1963)]{m63}
Marquardt, D.W. (1963).
An algorithm for the estimation of non-linear parameters.
{\em Journal of the Society for Industrial 
and Applied Mathematics\/}, 
{\bf 11}, 431--441.

\bibitem[Muller {\em et al\/}.(2025)]{metal25}
Muller, J., Mooney, K., Bowen, S.G., Klotzbach, P.J., Martin, T.,
Philp, T.J., Bhatt, D., Dixon, R.S.,  Girimurugang, S.B. (2025).
Normalized hurricane damage in the United States: 1900--2022.
{\em Bulletin of the American Meteorological Society\/},
{\bf 106}(1), E51--E67.

\bibitem[Pielke and Landsea(1998)]{pl98}
Pielke, Jr., R.A. and Landsea, C.W. (1998).
Normalized hurricane damages in the United States: 1925--1995.
{\em Weather and Forecasting\/}, {\bf 13}, 621--631.

\bibitem[Poudyal(2021a,b)]{p21a}
Poudyal, C. (2021a).
Robust estimation of loss models for lognormal 
insurance payment severity data.
{\em ASTIN Bulletin\/}, {\bf 51}(2), 475--507.

\bibitem[Poudyal(2021b)]{p21b}
Poudyal, C. (2021b).
Truncated, censored, and actuarial payment-type moments for 
robust fitting of a single-parameter Pareto distribution.
{\em Journal of Computational and Applied Mathematics\/}, 
{\bf 388}(May 2021), 113310.

\bibitem[Ruppert and Wand(1994)]{rw94}
Ruppert, D. and Wand, M.P. (1994). 
Multivariate locally weighted least squares regression.
{\em The Annals of Statistics\/}, 
{\bf 22}(3), 1346--1370.

\bibitem[Serfling(2002a)]{s02a} 
Serfling, R.J. (2002a).
{\em Approximation Theorems of Mathematical Statistics\/}.
Wiley.

\bibitem[Serfling(2002b)]{s02b}
Serfling, R. (2002b).
Efficient and robust fitting of lognormal distributions
(with discussion).
{\em North American Actuarial Journal\/},
{\bf 6}(4), 95--109.
Discussion: {\bf 7}(3), 112--116.
Reply: {\bf 7}(3), 116.

\bibitem[Tukey(1977)]{t77} 
Tukey, J.W. (1977).
{\em Exploratory Data Analysis\/}.
Addison Wesley.

\bibitem[Xu {\em et al\/}.(2014)]{xic14}
Xu, Y., Iglewicz, B., and Chervoneva, I. (2014).
Robust estimation of the parameters of $g$-{\em and\/}-$h$ 
distributions, with applications to outlier detection.
{\em Computational Statistics and Data Analysis\/}, 
{\bf 75}(July 2014), 66--80.

\bibitem[Zhao {\em et al\/}.(2018)]{zbg18}
Zhao, Q., Brazauskas, V., and Ghorai, J. (2018).
Robust and efficient fitting of severity models
and the method of Winsorized moments.
{\em ASTIN Bulletin\/}, {\bf 48}(1), 275--309.

\bibitem[Zhao and Poudyal(2024)]{zp24}
Zhao, Q. and Poudyal, C. (2024).
Credibility theory based on winsorizing.
{\em European Actuarial Journal\/}, 
{\bf 14}, 777--807.
\end{thebibliography}
\end{document}